\documentclass[pra, twocolumn,nofootinbib]{revtex4-1}

\usepackage{usepkg}
\usepackage{math}

\usepackage{sidecap}

\definecolor{darkred}{RGB}{200, 0, 0}
\definecolor{darkblue}{RGB}{0, 0, 180}
\definecolor{darkgreen}{RGB}{50, 150, 0}

\definecolor{col2}{HTML}{64A857}
\definecolor{col3}{HTML}{D1603D}
\definecolor{orange}{rgb}{1,0.5,0}

\newcommand*{\freq}[1]{\mathsf{freq}(#1)}

\newcommand*{\extra}[1]{}

\begin{document}

\title{Fully device independent Conference Key Agreement}

\author{J\'{e}r\'{e}my Ribeiro}
\affiliation{QuTech, Delft University of Technology, Lorentzweg 1, 2628 CJ Delft, The Netherlands}

\author{Gl\'{a}ucia Murta}
\affiliation{QuTech, Delft University of Technology, Lorentzweg 1, 2628 CJ Delft, The Netherlands}

\author{Stephanie Wehner }
\affiliation{QuTech, Delft University of Technology, Lorentzweg 1, 2628 CJ Delft, The Netherlands}

\begin{abstract}
  We present the first security analysis of conference key agreement (CKA) in the most adversarial model of device independence (DI).
  Our protocol can be implemented {by any experimental setup} that is capable of
  performing Bell tests (specifically, we introduce the ``Parity-CHSH'' inequality),
  and security can in principle be obtained for any violation of the Parity-CHSH
  inequality.
  {We use a direct connection between the $N$-partite
  Parity-CHSH inequality and the CHSH inequality. Namely the Parity-CHSH inequality can be
  considered as a CHSH inequality or another CHSH inequality (equivalent up to relabelling) depending on
  the parity of the output of $N-2$ of the parties.}
  We compare the asymptotic key rate for
  DICKA  to the case where the parties use $N-1$ DIQKD protocols in order to generate a common key. We show that for some
  regime of noise the DICKA protocol leads to better rates.
\end{abstract}

\maketitle

Quantum communication allows cryptographic security that is provably impossible to obtain using any classical means.
Probably the most famous example of a quantum advantage is quantum key distribution (QKD)~\cite{ekert91,bb84}, which allows two parties, Alice and Bob, to
exchange an encryption key whose security is guaranteed even if the adversary has an arbitrarily powerful quantum computer.
What's more, properties of entanglement lead to the remarkable feature that security is sometimes
possible
even if the quantum devices used to execute the protocol are largely untrusted. Specifically, the notion
of \emph{device independent (DI)} security~\cite{mayers98, BHK05, Acin07} models quantum devices as black boxes in which we may only choose measurement settings and observe measurement outcomes. Yet, the quantum state and measurements employed by such boxes are unknown,
and may even be prepared arbitrarily by the adversary.

Significant efforts have been undertaken to establish the security of device independent QKD~\cite{Acin07, PAB09, MPA10, MS14a, MS14, VV14, DIEAT}, leading to ever more sophisticated security proofs. Initial proofs assumed a simple model in which the devices act independently and identically (i.i.d.)~in
each round of the protocol. This significantly simplifies the security analysis since the underlying properties of the devices may first be estimated by gaining statistical confidence from the
observation of the measurement outcomes in the tested rounds. The main challenge overcome by the more recent security proofs~\cite{MS14a, MS14, VV14, DIEAT} was to establish security even if the devices behave arbitrarily from one round to the next, including having an arbitrary memory of the past that they might use to thwart the efforts of Alice and Bob. Assuming that the devices carry at least some memory of past interactions is an extremely realistic assumption due to technical limitations,
even if Alice and Bob prepare their own trusted, but imperfect, devices, highlighting the extreme importance of such analyses for the implementation of device independent QKD. In contrast, relatively little is known about device independence outside the realm of QKD~\cite{RPKHW16,KW16,Silman11,Kent2011,TFKW13}.

Conference key agreement \cite{BD95,CH-K05,QCKA} (CKA or N-CKA) is the task of distributing a secret key among $N$ parties. In order to achieve this goal, one could make use of $N -1$ individual QKD protocols to distribute $N - 1$ different keys between one of the parties (Alice) and the others ($\text{Bob}_1,\ldots,\text{Bob}_{N-1}$), followed by Alice using these keys to encrypt a common key to all the participants.
However the existence of genuine multipartite quantum correlations can bring some advantage to multipartite tasks, and, as shown in Ref.~\cite{QCKA}, exploring properties of genuine multipartite entanglement can lead to protocols with better performance for  conference key agreement.

Here we present the first security analysis of conference key agreement in the most adversarial model of device independence.
Our protocol can be implemented using any experimental setup that is capable of testing  the  Parity-CHSH inequality.
The Parity-CHSH inequality is an inequality we introduce here, that is closely related to the CHSH inequality and that extends it to $N$-parties.
Our proof is based on the relation between the $N$-partite Parity-CHSH inequality and the CHSH inequality~\cite{CHSH69}.
We also compare the asymptotic rates obtained for DICKA with the implementation of $N-1$ independent DIQKD, and show that for some regime of noise it is advantageous to perform DICKA.
The manuscript is organised as follows: In the next Section we present the protocol and state the security definitions for conference key agreement. Then we sketch the security proof of our DICKA protocol. We finish with a comparison of the asymptotic key rates. 
An expanded and detailed derivation of the security proof and the noise model for the asymptotic key rates are presented in the Appendix.

\section{The Protocol}

For a device independent implementation of CKA, we consider a protocol with $N$ parties: Alice who possesses one device
with two inputs $\{0,1\}$, and Bob$_1$ possesses a device with three inputs $\{0,1,2\}$, an Bob$_2$,$\ldots$,Bob$_{N-1}$ possess each a device with two inputs $\{0,1\}$. Every device has two outputs.
During the protocol, Alice and the Bobs randomly choose some rounds to test for the violation of the Parity-CHSH inequality. They abort the protocol if the frequency of rounds where they win the Parity-CHSH game do not reach a specified threshold $\delta$.
We also consider that Alice has a source for generation of the states, which is independent of her measurement device.

\begin{widetext}
\begin{framed}
\begin{Ptol}[DICKA]
   \label{Ptol:DICKA}   \hfill
    \begin{enumerate}
      \item For every round $i\in [n]$ do:
        \begin{enumerate}
          \item Alice uses her source to produce and distribute an $N$-partite state, $\rho_{A_i B_{(1\ldots N-1),i}}$, shared among herself and the $N-1$ Bobs.
          \item Alice randomly picks  $T_i$, s.t. $P(T_i=1)=\mu$, and publicly communicates it to all the Bobs. 
          \item If $T_i=0$ Alice and the Bobs choose $(X_i,Y_{(1...N-1),i})=(0,2,0,...,0)$,
          and if $T_i=1$ Alice chooses $X_i\in_R\{0,1\}$ uniformly at random, Bob$_1$
          chooses $Y_{(1),i}\in_R\{0,1\}$ uniformly at random, and Bob$_2$,\ldots,Bob$_{N-1}$ choose $(Y_{(2...N-1),i}) =(1,\ldots,1)$.
          \item Alice and the Bobs input the previously chosen values in their respective device and record the outputs as $ A_i', B'_{(1\ldots N-1),i}$.
        \end{enumerate}
      \item They all communicate publicly the list of bases $X_1^n {Y_{(1\ldots N-1)}}_1^n$ they used. \label{Pt_Ptol:comm_round}
      \item \textbf{Error correction:} \label{Error_correction} Alice and the Bobs apply an error correction protocol.
   We call $O_A$ the classical information that Alice sends to the
      Bobs. For the purpose of parameter estimation, the Bobs also send some error correction information for the bits produced during the test rounds ($T_i=1$), we denote $O_{(k)}$ the error correction information sent by Bob$_k$.
      If the error correction protocol aborts for at least one Bob then they abort the protocol. If it does not abort they obtain the raw keys $\tilde K_A=A', \tilde K_{B_{(1\ldots N-1)}}$.
      \item \textbf{Parameter estimation:} If $T_i=1$, Alice uses $A_i'$ and her
      guess on ${B'}_{(1...N-1),i}$ to set $C_i=1$ if they have won the N-partite
      Parity-CHSH game, and to set $C_i=0$ if they have lost it. If $T_i=0$, she sets
      $C_i=\bot$. She aborts if $\sum_i C_i<\delta \cdot \sum_i T_i$, where $\delta\in]p_{\min},p_{\max}[$.
      \item \textbf{Privacy amplification:} Alice and the Bobs apply a privacy amplification protocol  to
      create final keys $K_A,K_{B_{(1\ldots N-1)}}$. We denote $S$ the classical information publicly sent by Alice during this step.
          \end{enumerate}
\end{Ptol}
\end{framed}
\end{widetext}

%

\textit{Security Definitions.} {For completeness, before stating our main result, which establishes the secret key length of Protocol~\ref{Ptol:DICKA}},
 we first formalise what it means for a DICKA protocol to be secure.
As for QKD \cite{RennerThesis,PR14} the security of conference key agreement \cite{QCKA} can be split into two terms: \emph{correctness} and \emph{secrecy}. Correctness is a statement about how sure we are that the $N$ parties share identical keys, and secrecy is a statement about how much information the adversary can have about Alice's key.

\begin{Def}(Correctness and secrecy)\label{def:correct-secret}
  A DICKA protocol is $\epsilon_{\rm corr}$-correct  if Alice's and Bobs' keys,
  $K_A$, $K_{B_{(1)}},\ldots, K_{B_{(N-1)}}$, are all identical with probability at least $1-\epsilon_{\rm corr}$.
And it is $\epsilon_{\rm sec}$-secret, if Alice's key $K_A$ is
  $\epsilon_{\rm sec}$-close to a key that Eve is ignorant about. This condition can be formalized as
  \[p_{\hat \Omega} \cdot \left\|\rho_{K_A E {|\hat \Omega}} - \frac{\id_A}{2^l}\otimes \rho_{E {|\hat \Omega}}\right\|_{\tr}\leq \epsilon_{\rm sec},\]
  where $\| \cdot \|_{tr}$ denotes the trace norm, $l$ is the key length,  $\hat \Omega$ is the event of the protocol not aborting, and $p_{\hat \Omega}$ is the probability for $\hat \Omega$.

  If a protocol is $\epsilon_{\rm corr}$-correct and $\epsilon_{\rm sec}$-secret then it is $\epsilon^s$-correct-and-secret for any
$\epsilon^s\geq\epsilon_{\rm corr}+\epsilon_{\rm sec}$.
\end{Def}

So in general when we say that a CKA (or a QKD) protocol is $\epsilon^s$ secure, we mean that for any possible physical implementation of the protocol, either it aborts with probability higher than $1-\epsilon^s$ or it is $\epsilon^s$-correct-and-secret, according to Definition~\ref{def:correct-secret} (see Appendix section \ref{App:DICKA}).

A combination of Definition~\ref{def:correct-secret} and the Leftover Hashing Lemma~\cite{RennerThesis} relates the length of a secret key, that can be obtained from a particular protocol, with the smooth min-entropy of Alice's raw key $A'$ conditioned on Eve's information (see~\cite{RennerThesis} for a detailed derivation of this statement): An $\epsilon_{sec}$-secret key of size
\begin{align}
l=H_{\min}^{\epsilon}(A'|E)-2\log{\frac{1}{\epsilon_{PA}}}
\end{align}
can be obtained, for $\epsilon_{sec}>2\epsilon +\epsilon_{PA}$. {The conditional smooth min-entropy is defined as $H_{\min}^{\epsilon}(A|E)_{\rho}:=\sup_{\sigma \in \mathcal{B}(\rho)} H_{\min}(A|E)_{\sigma}$, with the supremum taken over all positive semi-definite operators $\epsilon$-close to $\rho$ in the purified distance (see ~\cite{TomQuantum}). For $A$ a classical register and $\sigma$ a quantum state, $H_{\min}(A|E)_{\sigma}$ represents the maximum probability with which Eve can guess the value of $A$ if they share the state $\sigma$. In general $ H_{\min}(A|E)_{\sigma}:=\sup_{\tau_E} \sup \{ \lambda : \sigma_{AE} \leq 2^{-\lambda} \id_A \otimes \tau_E\}$, where the supremum is taken over all quantum states $\tau_E$.}

Definition~\ref{def:correct-secret} was proved to be a criteria for composable security for QKD in the device dependent scenario \cite{PR14}.
However it is important to note that for the DI case it is not known whether such a criteria is enough for composable security. Indeed, Ref.~\cite{noncomposableDIQKD}
suggests that this is not the case if the same devices are used for generation of a subsequent key, since this new key can leak information about the first key. Following Ref.~\cite{DIEAT} we chose to adopt these definitions as the security criteria for DICKA.

Our main result establishes the length of a secure key that can be obtained from Protocol~\ref{Ptol:DICKA}.

\begin{Thm}\label{thm.keyrate}
Protocol~\ref{Ptol:DICKA} generates an  $\epsilon^s$-correct-and-secret key, with $\epsilon^s\leq\epsilon_{\rm PA} +2(N-1)\epsilon'_{\rm EC} +2 \epsilon+\epsilon_{\rm EA}$, of length:
 \begin{align}\label{eq:key_length}
     l= & \max_{p_{\min} \leq \delta_{\rm opt} \leq p_{\max}} \big( ({f}(\delta,\delta_{\rm opt})-\mu)\cdot n - \tilde v \sqrt{n}\big) \nonumber \\
     & \;\;\;+ 3\log(1-\sqrt{1-(\epsilon/4)^2})-2 \log(\epsilon_{\rm PA}^{-1}) \\
     & \;\;\; -{\rm leak_{EC}}(O_A)- \sum_{k=1}^{N-1} {\rm leak_{EC}}(O_{(k)}),\nonumber
   \end{align}
where $\epsilon'_{\rm EC}$ 
is an error parameter of the error correction protocol, $\epsilon_{\rm PA}$ is the privacy amplification error probability, $\epsilon_{\rm EA}$ is a chosen security parameter for the protocol, and $\epsilon$ is a smoothing parameter.
$\delta$ is the specified threshold bellow which the protocol aborts.
The function ${f}(\,\cdot\,,\delta_{\rm opt})$ is the tangent of $\hat f(\cdot)$ (see Eq.~\eqref{eq:min-tradeoff}) in the point $\delta_{\rm opt}$, where $\delta_{\rm opt} \in ] p_{\min}, p_{\max}[$ is a parameter to be optimized.
$\tilde v= 2\big(\log(13)+ (\hat f'(p_{\rm opt})/\mu+1) \big)\sqrt{1-2\log(\epsilon \cdot \epsilon_{\rm EA})}+2 \log(7) \sqrt{-\log(\epsilon_{\rm EA}^2(1-\sqrt{1-(\epsilon/4)^2}))}$.
And the leakages due to error correction, ${\rm leak_{EC}}$, can be estimated according to a particular implementation of the protocol.
\end{Thm}

The security proof of Protocol~\ref{Ptol:DICKA} consists  of two main steps: We first use the
recently developed Entropy Accumulation Theorem \cite{DFR16} to split the overall entropy of Alice's string, produced during the
protocol, into a sum of the entropy produced on each round of the protocol. Then we develop a new method to bound the entropy produced
in one round by a function of the violation of the $N$-partite Parity-CHSH inequality, which generalises the
bound for the bipartite case derived in~\cite{Acin07,PAB09}. {In the following Section we sketch the steps of the proof of Theorem~\ref{thm.keyrate}. An expanded and detailed derivation of this result is presented in the Appendix.}

\section{Security Analysis}

\noindent\textit{Step 1: Breaking the entropy round by round with the Entropy Accumulation Theorem (EAT).}
To prove the security of Protocol \ref{Ptol:DICKA} we need to lower bound the
smooth min-entropy of the string produced by Alice's device conditioned on all the information Eve obtains during the protocol (evaluated on the
output state of Protocol \ref{Ptol:DICKA} given the event $\hat \Omega$ of not aborting.),
\begin{align}\label{eq:entropy}
  \begin{split}
  H_{\min}^\epsilon \big({A'}_1^n| X_1^n {Y_{(1\ldots N-1)}}_1^n  T_1^n O_A O_{(1\ldots N-1)} E\big)_{\rho_{|\hat \Omega}},
\end{split}
\end{align}
where $E$ denotes Eve's quantum side information and all the other registers have been defined in Protocol \ref{Ptol:DICKA}. We can treat
the error correction information $O_A O_{(1\ldots N-1)}$that is communicated between Alice and the Bobs as  as a leakage:
\begin{align}\label{eq:leak}
  \begin{split}
  \eqref{eq:entropy} \geq & \,H_{\min}^\epsilon\big({A'}_1^n| X_1^n {Y_{(1\ldots N-1)}}_1^n  T_1^n E\big)_{\rho_{|\hat \Omega}}\\
  & \hspace{3mm}-{\rm leak_{EC}}(O_A)  - \sum_{k=1}^{N-1} {\rm leak_{EC}}(O_{(k)}).
\end{split}
\end{align}
This relation follows from the properties of the smooth min-entropy (see \cite[Lemma 6.8]{T15}).

Now, in order to bound the term $H_{\min}^\epsilon\big({A'}_1^n| X_1^n {Y_{(1\ldots N-1)}}_1^n  T_1^n E\big)_{\rho_{|\hat \Omega}}$,
we use the Entropy Accumulation Theorem (EAT) \cite{DFR16}. The EAT has already been used
to prove security of device independent QKD \cite{DIEAT}. This theorem permits to
lower bound the above entropy by a sum of Von Neumann entropies evaluated on
each round $i$. More precisely:
\begin{align}
  \begin{split}\label{eq:EAT}
  H_{\min}^\epsilon\big({A'}_1^n| X_1^n {Y_{(1\ldots N-1)}}_1^n & T_1^n E\big)_{\rho_{|\hat \Omega}} \\ &\geq n t - v \sqrt{n},
  \end{split}
\end{align}
where $v$ is a prefactor independent of the number of rounds and $t$ is a lower bound (for every round $i$) on the Von Neumann entropy $H(A_i'|X_1^i {Y_{(1\ldots N-1)}}_1^i {A'}_1^{i-1} T_1^i E)_{\mathcal{M}_i(\sigma)}$
for all initial states $\sigma$ that 
would achieve a Bell violation larger than the chosen threshold $\delta$ (see Appendix Section \ref{App:EAT}).
The EAT then reduces the security proof in the most adversarial scenario to the estimation of  $t$.\vspace{1em}


\noindent \textit{Step 2: Bounding the entropy by a function of the Bell violation.}
We now proceed to lower bound $t$ for Protocol~\ref{Ptol:DICKA}, \textit{i.e.} we find
a lower bound on the Von Neumann entropy $H(A_i'|X_1^i {Y_{(1\ldots N-1)}}_1^i {A'}_1^{i-1} T_1^i E)_{\mathcal{M}_i(\sigma)}$
as a function of the violation of the Parity-CHSH inequality
for $N$ parties. The Parity-CHSH inequality
is an $N$-partite Bell inequality that reduces to the CHSH inequality for $N=2$. \\
\textbf{The CHSH inequality} can be formulated as a bound on the winning
probability of the following bipartite game. Let Alice and Bob be the two players in this
game called the CHSH game. At the beginning of the game, they are both asked a uniformly random binary question
$x\in\{0,1\}$ and $y\in\{0,1\}$ respectively. They then have to answer bit $a$ and $b$
 respectively. They win the game if and only if \[a+b=xy \text{ mod }2.\]
 No communication is allowed between Alice and Bob during the game. They can, however, agree
 on any strategy before the start of the game. The CHSH inequality states that by using a classical strategy
 (a non-quantum strategy)\footnote{strategies that can be modeled with local hidden
 variables} Alice and Bob's winning probability must satisfy the following,
\begin{align}P_{\rm win}^{\rm CHSH} \leq \frac{3}{4}.\end{align}

By a small change in the CHSH inequality we can extend it to $N>2$ parties.

\begin{Def}[Parity-CHSH] The Parity-CHSH inequality extends the CHSH inequality to $N$ parties as follows. Let Alice, Bob$_1$, \ldots, Bob$_{N-1}$ be the $N$ players of the following game (the Parity-CHSH game).
Alice and Bob$_1$ are asked uniformly random binary questions $x\in\{0,1\}$ and $y\in\{0,1\}$ respectively. The other
Bobs are each asked a fixed question, \emph{e.g.}~always equal to $1$. Alice will answer bit $a$, and $\forall i \in \{1,\ldots,N-1\}$, Bob$_i$ answers bit $b_i$. We denote by $\bar b:=\bigoplus_{2 \leq i\leq N-1} b_i$,
the parity of all the answers of Bob$_2$,\ldots,Bob$_{N-1}$. The players win if and only if
\[a+b_1=x(y+\bar b) \text{ mod }2.\]
As for the CHSH inequality, classical strategies for the Partity-CHSH game must satisfy,
\begin{align}
    P_{\rm win}^{\rm Parity-CHSH} \leq \frac{3}{4}.
\end{align}

\end{Def}

\begin{Rmk}\label{Rmk:rmk2}
 Note that if we condition on $\bar b=0$, the game is essentially the CHSH game. When conditioned on $\bar b=1$ the
 Parity-CHSH game reduces to a game equivalent to the CHSH up to relabelling the question $y$. We will use this to later
 prove that the function $\hat f$ defined in eq.~\eqref{eq:min-tradeoff} lower bounds some entropy of interest.
\end{Rmk}

We can use the violation of this inequality to lower the von Neumann entropy $H:=H(A_i'|X_1^i {Y_{(1\ldots N-1)}}_1^i {A'}_1^{i-1} T_1^i E)_{\mathcal{M}_i(\sigma)}$.
We can do this by using that the Parity-CHSH inequality is very similar to the CHSH inequality
and that Ref.~\cite{Acin07} essentially lowerbounds $H$ as a function of the violation CHSH
inequality.

\begin{Thm}
  The function $\hat f$ defined below lower bounds $H$.
\begin{align}\label{eq:min-tradeoff}
    \!\!\!\!\!\hat f(p_w)\!:=\!\!\left(1-\frac{\mu}{2}\right)\!\!\left(1-h\left(\frac{1}{2} + \frac{1}{2}\sqrt{(4p_w-2)^2-1}\right)\! \right),
\end{align}
where $p_w$ is a shorthand notation for $P_{\rm win}^{\rm Parity-CHSH}$, and where the winning probability
is evaluated on the state $\mathcal{M}_i(\sigma)$ on which the entropy is evaluated.
\end{Thm}
\begin{proof}[Sketch of proof]
  We first notice that since $\Pr(X_i=0)=(1-\frac{\mu}{2})$,
  \begin{align*}
      H=&\Big(1-\frac{\mu}{2}\Big)H(A_i'|X_1^{i-1} {Y_{(1,\ldots,N-1)}}_1^i {A'}_1^{i-1} T_1^iE, X_i=0)&\\
      &+ \frac{\mu}{2} \underbrace{H(A_i'|X_1^{i-1} {Y_{(1,\ldots,N-1)}}_1^i {A'}_1^{i-1} T_1^iE, X_i=1)}_{\geq 0}&\\
      \geq&\Big(1-\frac{\mu}{2}\Big)H(A_i'|X_1^{i-1} {Y_{(1,\ldots,N-1)}}_1^i {A'}_1^{i-1} T_1^iE, X_i=0).
  \end{align*}
  The above inequality holds since $A_i'$ is a classical register.
  Conditioned on $X_i=0$, $A'_i$ is independent of ${Y_{(1,\ldots,N-1)}}_i$ and of
  $T_i$, and in the following, $R$ denotes the registers $X_1^{i-1} {Y_{(1,\ldots,N-1)}}_1^{i-1} {A'}_1^{i-1} T_1^{i-1} E$ so that,
  \begin{align*}
      &H(A_i'|X_1^{i-1} {Y_{(1,\ldots,N-1)}}_1^i {A'}_1^{i-1} T_1^iE\, X_i=0)\\
      &=H(A_i'|R\, X_i=0).
  \end{align*}
  It remains to lower bound $H(A_i'|R, X_i=0)$. We first lower bound it by
  \begin{align*}
      H(A_i'|R\, X_i=0) \geq H(A_i'|R, X_i=0,\, \bar b),
  \end{align*}
  where $\bar b$ is the register that contains the parity bit of the outcome of
  Bob$_2$,\ldots,Bob$_{N-1}$.
  We can then expand the Von Neumann entropy as,
  \begin{align*}
      H(A_i'|R\, X_i=0,\, \bar b)=&p_{\bar b=0} H(A_i'|R, X_i=0,\, \bar b=0)\\
      &+ p_{\bar b=1}H(A_i'|R, X_i=0,\, \bar b=1).
  \end{align*}
  From \cite{Acin07} we have that $(1\!-\!\frac{\mu}{2}) H(A_i'|R, X_i=0, \bar b=0) \geq  \hat f(p_{w|\bar b=0})$ and
  $(1-\frac{\mu}{2}) H(A_i'|R, X_i=0,\, \bar b=1)\geq  \hat f(p_{w|\bar b=1})$. Indeed, from Remark \ref{Rmk:rmk2} we have that conditioned on $\bar b=0$ the Parity-CHSH game is simply a CHSH game, therefore $p_{w|\bar b=0}$
  is equal to $P_{\rm win}^{\rm CHSH}$ when evaluated on the state shared between Alice and Bob$_1$ conditioned on $\bar b=0$. Moreover  \cite{Acin07}
  precisely lower bounds $(1-\frac{\mu}{2})H(A_i'|R\, X_i=0,\, \bar b=0)$ by $\hat f\big(P_{\rm win}^{\rm CHSH}\big)$. The same reasoning holds for $\bar b=1$.

  As a consequence,
  \begin{align*}
      \Big(1\!-\!\frac{\mu}{2}\Big) H(A_i'|&R\, X_i=0,\, \bar b)\\ &\geq p_{\bar b=0}\hat f(p_{w|\bar b=0})+ p_{\bar b=1} \hat f(p_{w|\bar b=1}).
  \end{align*}
  By convexity of the function $\hat f$, we get,
  \begin{align*}
      \Big(1\!-\!\frac{\mu}{2}\Big) H(A_i'|R\, X_i=0,\, \bar b)\geq&\hat f\big(p_{\bar b=0} p_{w|\bar b=0}+ p_{\bar b=1} \hat p_{w|\bar b=1}\big)\\
      =& \hat f\big(p_w\big),
  \end{align*}
  and therefore $H\geq \hat f(p_w).$
\end{proof}

\section{Asymptotic key rate and comparison with DIQKD based protocol}

\begin{figure}[h]
    \includegraphics[width=260pt]{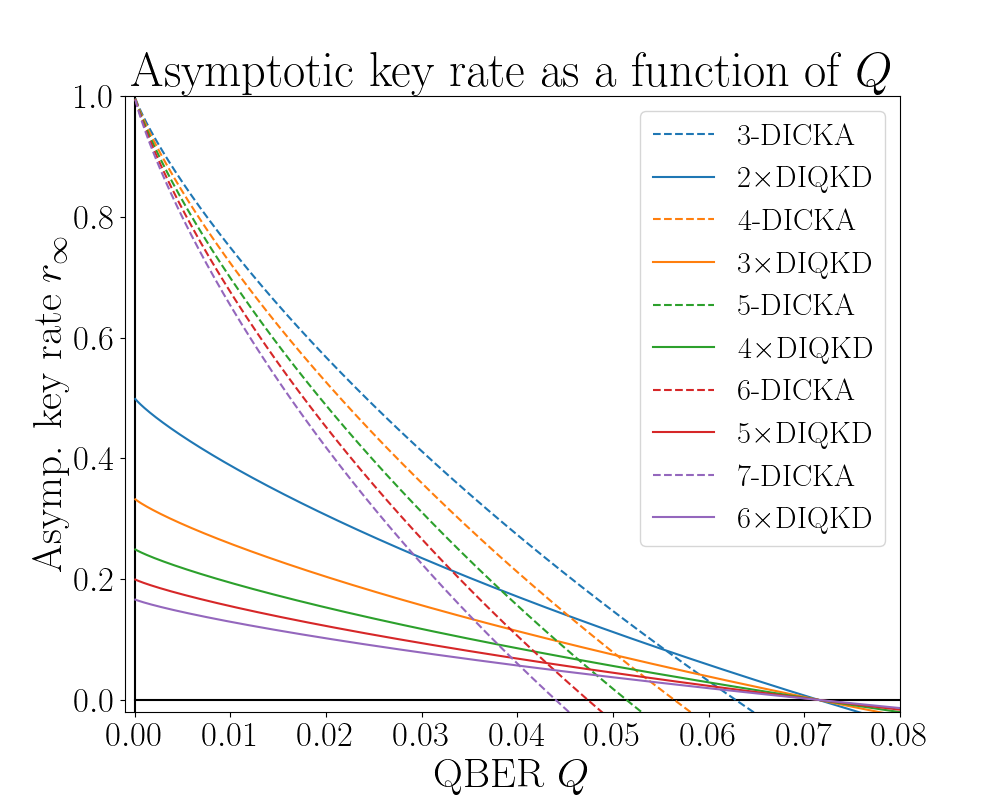}
    \caption{Asymptotic key rate for 
    $N$-DICKA (dashed lines), and for the distribution of a secret key between $N$ parties through $N-1$ DIQKD protocols (solid lines), when each qubit experiences independent bit errors  measured at a bit error rate (QBER) $Q$.
    From top to  bottom, the lines correspond to $N=\{3,4,5,6,7\}$.
    We observe that for low noise regime it is advantageous to use DICKA  instead of $(N-1)\times$DIQKD \cite{DIEAT}. In general, the comparison between the two methods depends on the cost and noisiness of producing GHZ states over pairwise EPR pairs.}
    \label{fig:asymp_key_rate1}
  \end{figure}

{We remark that bipartite QKD has of course been studied in the device independent setting \cite{DIEAT}, but as we are going to see in Figure \ref{fig:asymp_key_rate1}, a conference key agreement protocol can be beneficial for certain regimes of noise.}

  Combining Eqs.~\eqref{eq:entropy},~\eqref{eq:leak} and~\eqref{eq:EAT} we get a lower bound on the length of secret key we can obtain with Protocol~\ref{Ptol:DICKA}, which, when divided by the number of rounds $n$,
  gives us a lower bound on the secret key rate.

  In order to calculate the secret key rate, we also need to estimate the leakages due to error correction,
  Eq.~\eqref{eq:leak}, and for that we need to specify the model for an honest implementation. Modeling  the noise
  on the distributed state as a depolarising noise we get:
  \begin{align}
    {\rm leak_{EC}}(O_A) \leq  ((1-\mu) h(Q)+\mu) n + \mathcal{O}(\sqrt{n}),
  \end{align}
  and
  \begin{align}
    {\rm leak_{EC}}(O_{(k)}) \leq \mu n + \mathcal{O}(\sqrt{n}),
  \end{align}
where
$Q$ is the \emph{quantum bit error rate} (QBER) between Alice and one of the Bobs.
A detailed calculation of the leakage for this particular noise model is presented in Appendix Section \ref{Sec.keyrate}.

Using this estimation of the leakage in the bounds for the entropy \eqref{eq:entropy},
and by taking $\mu \to 0$, s.t. $\mu \sqrt{n} \to \infty$, we get the asymptotic key rate for Protocol \ref{Ptol:DICKA}:
\begin{widetext}
 \begin{align}\label{rateQ}
     r_{N-\text{CKA},\infty}=1-h\left( \frac{1}{2}+ \frac{1}{2}\sqrt{16\left(\frac{\sqrt{1-2Q}^N}{2\sqrt{2}} + \frac{(1-2Q)\left(1-\sqrt{1-2Q}^{N-2}\right)}{4\sqrt{2}}\right)^2-1} \right)-h(Q).
 \end{align}
 \end{widetext}

We compare the above rate with the one we would have if  Alice was performing $N-1$ DIQKD protocols in order to establish a common key with all the Bobs \cite{DIEAT}:
  \begin{align}
    r_{_{{(N-1)\times{\rm DIQKD}}}}^{\infty}= \frac{1-h\left( \frac{1}{2}+ \frac{1}{2}\sqrt{{2 (1- 2Q)^2 }-1} \right)-h(Q)}{N-1}.
  \end{align}
   Because when Alice runs $N-1$ DIQKD protocols she needs $n$ rounds for
  each of the $N-1$ Bobs, the key rate $r_{_{(N-1)\times{\rm DIQKD}}}^{\infty}$ gets a factor of $\frac{1}{N-1}$. Note that here we consider that the cost for locally producing an $N$-partite GHZ state is comparable to the cost of producing EPR pairs. An analysis taking into account
  these costs for particular implementations will lead to a more fair comparison.

   A comparison of these key rate is given in Figure \ref{fig:asymp_key_rate1}, where we see
   that in some regime of noise, it can be advantageous to use the $N$-partite
   DICKA Protocol \ref{Ptol:DICKA} instead of $N-1$ independent
   DIQKD protocols.

  \section{Conclusion}
We presented the first security proof for a fully device independent implementation of conference key agreement.
We have shown that, in principle, security can be achieved for any violation of the  Parity-CHSH inequality.

We have compared the asymptotic key rates achieved with the DICKA protocol versus $N-1$ implementations of DIQKD, modelling the quantum channel connecting the parties as depolarising channels. 
For implementations where the cost of local generation of GHZ states and EPR pairs is comparable, we show that it is advantageous to use DICKA for low noise regimes.
A careful analysis that takes into account the costs of generation of the states is still needed for particular implementations.

{We remark that proving advantage for a small number of parties already leads to better protocols for networks. Indeed, instead of using DIQKD as building block for an $N$-DICKA protocol (for large $N$), one can use $k$-DICKA protocols,
upon availability of $k$-GHZ states for $k=3,4$ or $5$.}

{Finally, we also}
{remark that our DICKA protocol can be adapted for other multipartite Bell inequalities. However,}
in general, finding good lower bounds on Eve's information about Alice's output as a function of the Bell violation is a difficult task.

\section*{Acknowledgements}
We thank Johannes Borregaard, Mikhail D. Lukin and  Valentina Caprara Vivoli for helpful discussions.
This work was supported by  STW Netherlands, NWO VIDI and ERC Starting Grant. We also thank Aby Philip and Eneet Kaur for spotting a mistake in Eq.~\eqref{pwin} (and as a consequence in Eq.~\eqref{rateQ}). The extra factor of 2 in the last term of~\eqref{pwin} is corrected in this arXiv version.\\

This updated version corrects a mistake we have made in the previous version.
This mistake is discussed in [arXiv:1906.01710].\\
In order to be clear for the reader we give here some explanations.
In the former version of the article we present a protocol for device-independent
conference key agreement (CKA) between $N$ parties, Alice, Bob$_1$,\ldots,
Bob$_{N-1}$, using an $N$-partite GHZ state ($(\ket{0}^{\otimes N}+\ket{1}^{\otimes N})/\sqrt{2}$).
The protocol, aiming to be
secure in the device-independent settings, relies on a statistical Bell test.
In particular, in the former version of the article we present the $N$-partite Mermin-Ardehali-Belinskii-Klyshko
(MABK) inequality. However, using this inequality with the GHZ state, leads to a protocol
that is secure but does not produce key. The intuition for that is the following.
\begin{enumerate}
\item In order to ensure security,
the protocol requires that the state and the measurement are such that they can achieve
a sufficiently high violation of the MABK inequality. To do so using the GHZ state,
Alice's observables $A_0$ and $A_1$ need to be in the $XY$ plane of the Bloch sphere.
\item In order to generate key that is correlated with the Bobs, Alice needs to have at
least one of her observables (either $A_0$ or $A_1$) that is equal to the Pauli $Z$ operator.
\end{enumerate}
The two above conditions about $A_0$ and $A_1$ cannot simultaneously be true.

Moreover even if there is no noise in the protocol, if Alice measures the GHZ
state with a measurement in the XY plane, her outcomes will be completely uncorrelated
with Bobs' outcomes. Therefore no key can be produced, even though violation of the
MABK inequality ensures that Alice outcomes have high entropy conditioned on Eve.
As a consequence, the protocol of the former version of the article will abort at almost every honest
execution, and hence no key is produced.
Of course one could consider
measuring the GHZ state in a basis in between the $Z$ basis and the $XY$ plane.
However,
this would  at best lead to very low key rate, and it would at worst not be
sufficient to get any key at all, causing the protocol to always abort.

We solve this issue by changing the inequality we use in the protocol.
The new inequality (the Parity-CHSH inequality) we introduce in this update is
such that violation can be achieved by measuring in the $Z$ basis, which ensures
that entropy conditioned on Eve of Alice's measurement outcomes in the $Z$ basis
is high. Furthermore, when all the parties measure the GHZ state in the $Z$ basis
they should get the same outcome (in the noiseless scenario), which allows for the
production of a shared bit string (the key). If a small amount of noise is present,
the errors it induces can be corrected by an error correction procedure as already
presented in the protocol of the former version of the article.

\newpage

\appendix*
\setcounter{equation}{0}

\begin{widetext}
  \section*{Appendix}\label{Appendix}
Here we expand in detail upon the security proof of the device independent conference key agreement protocol (DICKA) presented in the main text,
Protocol \ref{Ptol:DICKA}. A more detailed version of Protocol \ref{Ptol:DICKA} is given in this appendix in
Protocol \ref{PtolA:DICKA}.

The Appendix is organized as follows: In Section~\ref{Sec.preliminaries} we introduce
some background. We start by introducing the notation and some definitions which are going to be used in the main proofs.
Then we present the entropy accumulation theorem (EAT), which constitutes an important tool of our security proof.
We finish discussing the set of hypothesis contained in the device independent (DI) model. 
In Section~\ref{Sec.proof}, we state the DICKA protocol and present the detailed security proof.
In Section~\ref{Sec.keyrate} we present the noise model to compare the asymptotic key rate of the DICKA protocol to
the case where the parties perform $N-1$ independent DIQKD protocols in order to generate a common key.


\subsection{Preliminaries}\label{Sec.preliminaries}

\subsubsection{Notation}
We denote $\mathcal{H}_A$ the Hilbert space of the system $A$ with
dimension $|A|$ and $\mathcal{H}_{AB}:= \mathcal{H}_A \otimes
\mathcal{H}_B$ the Hilbert space of the composite system, with $\otimes$ the tensor product.
By $\mathcal{L}(\mathcal{H})$, $\mathcal{S}_a(\mathcal{H})$, $\mathcal{P}(\mathcal{H})$ and $\mathcal{S}(\mathcal{H})$ we mean the set of linear,
self-adjoint, positive semidefinite and (quantum) density operators on $\mathcal{H}$, respectively. For two operators $A,B \in
\mathcal{S}_a(\mathcal{H})$, $A \geq B$ means $(A-B) \in \mathcal{P}(\mathcal{H})$. For $M \in \mathcal{L}(\mathcal{H})$, we denote
$|M|:=\sqrt{M^\dagger M}$, and the Schatten $p$-norm $\|M\|_p:=\tr(|M|^p)^{1/p}$ for $p \in [1,\infty[$, and $\|M\|_\infty$ is the largest singular value
of $M$. For $M\in \mathcal{P}(\mathcal{H})$, $M^{-1}$ is the generalised
inverse of $M$, meaning that the relation $MM^{-1}M=M$ holds.
If $\rho_{AB} \in \mathcal{S}(\mathcal{H}_{AB})$ then we denote $\rho_A:=\tr_B(\rho_{AB})$ and $\rho_B:=\tr_A(\rho_{AB})$ to be the respective
reduced states. We use $[n]$ as a shorthand for $\{1,\dots,n\}$. If we deal with a system composed with $N$ subsystems within a round $i$ of a protocol we denote
$A_{(k\ldots l),i}$ for $A_{(k),i},\ldots A_{(l),i}$ ($k,l\in [N]: k\leq l$), where $A_{(k),i}$ is the $k^{\rm th}$ subsystem of the round $i$. If we deal with a
system composed of $n$ subsystems across the $n$ rounds of a protocol we denote
$A_k^l$ for $A_k,\ldots,A_l$ ($k,l\in [n]: k\leq l$). Therefore ${A_{(k\ldots l)}}_m^o$ is a short for $A_{(k\ldots l),m},\ldots A_{(k\ldots l),o}$ ($k,l\in [N],m,o\in [n]: k\leq l, m\leq o$).

For classical-quantum states (or cq-states)
\begin{align*}
    \rho_{XA}:=\sum_{x\in \mathcal{X}} p_x \cdot \ketbra{x}{x}_X\otimes \rho_{A|x},
\end{align*}
where $\{p_x\}$ is a probability distribution on the alphabet $\mathcal{X}$ of $X$. We define a cq-state $\rho_{XA|\Omega}$ conditioned on an event $\Omega \subset \mathcal{X}$ as,
\begin{align}
    \rho_{XA|\Omega}:= \frac{1}{p_\Omega} \sum_{x\in \Omega} p_x \cdot \ketbra{x}{x}_X\otimes \rho_{A|x},\ \mathrm{where}\ p_\Omega:=\sum_{x \in \Omega} p_x. \label{eq:cond_state}
\end{align}

We will denote by CPTP maps the linear maps that are Completely Positive and Trace Preserving.

Let $\mathcal{C}$ be an alphabet, and $C_1, \ldots, C_n$ be $n$ random variables on this alphabet. We call $\freq{C_1^n}$ the vector whose
components labeled by $c\in \mathcal{C}$ are the frequencies of the symbol $c$:
\[\freq{C_1^n}_c := \frac{|\{i: C_i=c\}|}{n}.\]

\subsubsection{Entropies}

Throughout this work we will make use the smooth min- (max-) entropy. To define them we first define the min- and max-entropies \cite{T15}.
\begin{Def}
  If $\rho_{AB}$ is a bipartite state and $\epsilon \in ]0,1[$, we define the min- and max-entropies as,
  \begin{align}
    H_{\min}(A|B)_{\rho} &:=-\log\left(\inf_{\sigma_B} \| \rho_{AB}^{\frac{1}{2}} \sigma_B^{-\frac{1}{2}}\|_\infty^2 \right)\\
       H_{\max}(A|B)_{\rho}&:=\log\left(\sup_{\sigma_B} \| \rho_{AB}^{\frac{1}{2}} \sigma_B^{-\frac{1}{2}}\|_1^2 \right),
  \end{align}
  where the infimum and the supremum are taken over all states $\sigma_B \in \mathcal{S}(\mathcal{B})$.
  Their smooth versions are defined as
  \begin{align}
    H_{\min}^{\epsilon}(A|B)_{\rho} &:= \sup_{\hat \rho_{AB}} H_{\min}(A|B)_{\hat \rho}\\
      H_{\max}^{\epsilon}(A|B)_{\rho}&:=\inf_{\hat \rho_{AB}} H_{\max}(A|B)_{\hat \rho},
  \end{align}
  where the supremum and infimum are over all operators $\hat \rho_{AB} \in \mathcal{P}(\mathcal{H}_{AB})$ in a $\epsilon$-ball
  (in the purified distance) centered in $\rho_{AB}$. Moreover if $A$ is classical, the optimization can be restricted to
  an $\epsilon$-ball in $\mathcal{S}(\mathcal{H}_{AB})$.
\end{Def}

\subsubsection{Markov Condition}
The technique we are going to use for the security analysis of our DICKA protocol strongly relies on the fact that some variables satisfy the
 so-called  \emph{Markov Condition}.
\begin{Def}[Markov Condition]
  Let $\rho_{ABC}$ be a state in $\mathcal{S}(\mathcal{H}_{ABC})$. We say that $\rho_{ABC}$ satisfies the Markov
  condition $A \leftrightarrow B \leftrightarrow C$ if and only if
  \begin{align}
      I(A:C|B)_\rho=0,
  \end{align}
  where $I(A:C|B)_\rho$ is the mutual information between $A$ and $C$ conditioned on $B$ for the state $\rho_{ABC}$.
\end{Def}
This condition becomes trivial when $A$ $B$ and $C$ are independent random variables. For more details on the definition of the
 \emph{Markov condition} see \cite[section 2.2 \& appendix C]{DFR16}.

\subsubsection{The Entropy Accumulation Theorem (EAT)}\label{App:EAT}

The security proof of our DICKA protocol makes use of a very powerful tool called Entropy Accumulation Theorem (EAT), recently introduced in \cite{DFR16}. The EAT relates
the smooth min- (max-) entropy of $N$ subsystems to the Von Neumann entropy of each subsystem. In this section we recall some necessary
definitions from \cite{DFR16} and state the EAT.

The entropy accumulation theorem applies to states of the form,
\begin{align}
  \rho_{C_1^nA_1^nB_1^nE} :=(\tr_{R_n} \circ \mathcal{M}_n\circ \ldots \circ \mathcal{M}_1 \otimes \id_E)(\rho_{R_0E}),
  \label{def:rho}
\end{align}
for some initial state $\rho_{R_0E}\in \mathcal{S}(\mathcal{H}_{R_0 E})$ and, $\forall i\in [n]$, $\mathcal{M}_i$ is a
\emph{EAT channel} defined as follows.
\begin{Def}[EAT channels (from \cite{ADFORV18})]
  For $i\in[n]$ we call $\mathcal{M}_i$ a \emph{EAT channel} if $\mathcal{M}_i$ is a CPTP map from $R_{n-1}$ to $C_i A_i B_i R_i$
  such that $\forall i \in [n]$:
  \begin{enumerate}
    \item $A_i,B_i,C_i$ are finite dimensional systems, $C_i$ is classical and $R_i$ is an arbitrary quantum system.
    \item For any state $\sigma_{R_{i-1}R}$, where $R$ is isomorphic to $R_{i-1}$, the output state $\sigma_{R_i A_i B_i C_i R}:=
    (\mathcal{M}_i\otimes \id_R)\sigma_{R_{i-1}R}$
    is such that the classical register $C_i$ can be measured from $\sigma_{A_i B_i}$.
    \item Any state defined as in \eqref{def:rho} satisfies the following Markov conditions,
    \begin{align}
      \forall i \in [n],\ A_1^{i-1} \leftrightarrow B_1^{i-1} E \leftrightarrow B_i .
      \label{eq:Markov}
    \end{align}
  \end{enumerate}
\end{Def}

To state EAT we also need the notion of min- and max-tradeoff functions. Let $\mathds{P}(\mathcal{C})$ be the set of distributions
on the alphabet $\mathcal{C}$ of $C_i$. For any $q \in \mathds{P}(\mathcal{C})$ we define the set of states
\begin{align}
  \Sigma_i(q):= \{\sigma_{C_i A_i B_i R_i R}= (\mathcal{M}_i\otimes \id_R)(\sigma_{R_{i-1} R}): \sigma_{R_{i-1} R} \in \mathcal{S}(\mathcal{H}_{R_{i-1}R})\ \&\ \sigma_{C_i}=q\}.
\end{align}

\begin{Def}
  A real function f on $\mathds{P}(\mathcal{C})$ is called a \emph{min-tradeoff} function for a map $\mathcal{M}_i$ if
  \begin{align}
    f_i(q)\leq \inf_{\sigma \in \Sigma_i(q)} H(A_i|B_i R)_\sigma,
  \end{align}
  and \emph{max-tradeoff} function for a map $\mathcal{M}_i$ if
   \begin{align}
    f_i(q)\geq \sup_{\sigma \in \Sigma_i(q)} H(A_i|B_i R)_\sigma.
   \end{align}
   If $\Sigma_i(q)= \varnothing$, the infimum is taken to be $+\infty$ and the supremum $-\infty$.
   \label{def:tradeoff_fct}
\end{Def}

We can now state the EAT.
\begin{Thm}[EAT from \cite{DFR16}, Theorem 4.4]\hfill \\
  Let $\mathcal{M}_1, \ldots, \mathcal{M}_n$ be a EAT channel and $\rho_{C_1^n A_1^n B_1^n E}$ be a state as defined in \eqref{def:rho},
  let $h\in \mathbb{R}$, $f$ be an affine min-tradeoff function for all the maps $\mathcal{M}_i\,, i\in [n]$, and $\epsilon \in ]0,1[$.
  For any event $\Omega \subset \mathcal{C}^n$ such that $f(\freq{C_1^n})\geq h$,
  \begin{align}
        H_{\min}^{\epsilon}(A_1^n|B_1^nE)_{\rho_{|\Omega}} \geq n h -v \sqrt{n},
  \end{align}
  where $v=2(\log(1+2d_A)+\lceil \|\nabla f\|_\infty
\rceil)\sqrt{1-2\log(\epsilon \cdot p_\Omega)}$, where $d_A$ is the
maximum dimension of the system $A_i$. On the other hand we have,
  \begin{align}
    H_{\max}^{\epsilon}(A_1^n|B_1^nE)_{\rho_{|\Omega}} \leq n \tilde h +v \sqrt{n},
  \end{align}
  where we replace $f$ by an affine \emph{max-tradeoff function} $\tilde f$, such that the event $\Omega$ implies $\tilde h \geq \tilde f(\mathrm{freq}(C_1^n))$.
\end{Thm}



\subsubsection{Device Independent assumptions}

When dealing with cryptographic tasks it is important to be precise under which assumptions a protocol is proven secure. If an assumption is not satisfied in a particular implementation, the entire security of the protocol may be compromised. The device independent framework allows one to relax many strong assumptions about the underlying system and devices, however some assumptions (without which we can probably not achieve any security) are still present and it is important to make them explicit. In the following we state the assumptions present in our model, which constitutes the standard set of assumptions made in all device independent protocols. This minimal set of assumptions is crucial for security in the device independent framework, as a relaxation of any of
them compromises the security of the protocol.

\begin{Hyp}\label{Hyp:DICKA}
    Our DICKA protocol considers $N$ parties, namely Alice, $\text{Bob}_1,\ldots,\text{Bob}_{N-1}$, and the eavesdropper,
 Eve.
  They satisfy the following assumptions:
  \begin{enumerate}
    \item Each party is in a lab  which is isolated from the outside (in particular from Eve). As a consequence no non-intended information can go in or out of the labs.
    \item Each party holds a trusted random number generator (RNG).
    \item All classical communications between the parties are assumed to be authenticated, and all classical operations are assumed to be trusted.
    \item Each party has a measurement device in their lab in which they can input classical information and which outputs $0$ or $1$. The
    measurement devices are otherwise arbitrary, and therefore could be prepared by Eve.
    \item Alice has a source that produces some $N$ partite quantum state $\rho_{A_i B_{(1\ldots N-1),i}}$ in the round $i$. We allow Eve to hold the purification of $\rho_{A_1^n {B_{(1\ldots N-1)}}_1^n}$
    (the state between Alice and the Bobs for the $n$ rounds of the protocol)
    and we denote the pure global state $\rho_{A_1^n {B_{(1\ldots N-1)}}_1^n E}$. This source is also assumed to be arbitrary, and therefore we can assume that it is prepared by Eve.
    \item \label{iso_source_device} We will assume that Alice's source and her measurement device are independent ({\it e.g.}~Alice can isolate the source from the measurement device). Therefore there is no non-intended communication between the source and
    her measurement device.
  \end{enumerate}
\end{Hyp}

Point \ref{iso_source_device} of Assumptions \ref{Hyp:DICKA} is usually not explicitly
stated in  previous works on device independent
  QKD, however we remark that this assumption is also present in all previous protocols.
Indeed assumption \ref{iso_source_device} is important to guarantee that no extra information about the outcomes of Alice's device is leaked to Eve
  (since Alice and Bob are in isolated labs), apart from what
  she can learn from the purifying system in her possession and the classical communication intentionally leaked during the protocol.
Previous protocols usually assume that an external source is responsible for producing the states. However note that in order to distribute the states to
 Alice and Bob's devices one need a quantum channel connecting the external source with their labs, and similarly it is assumed that no information from the
 devices is leaked through this quantum channel.
 An alternative approach is to
 assume that the full state for the $n$ rounds of the protocol is already shared
  between the two parties at the very beginning of the protocol (and any quantum channel connecting the source and the devices is disconnected once the
  protocol starts). 
 However this is an unrealistic assumption, since an implementation of such protocol would require quantum memory to last for the entire duration of the
 protocol.
For that reason, here we chose NOT to assume that the
  state is already shared among all the parties, and assumption \ref{iso_source_device} prevents the simple attack
  described in \cite[Appendix C]{RMW16}, where the outcome of round $i$ is leaked throughout the state transmitted to Bob in the next rounds.

\subsection{From self-testing to Device Independent Conference Key Agreement}\label{Sec.proof}

The Clauser-Horne-Shimony-Holt (CHSH) inequality \cite{CHSH69} has been successfully used to prove security of DIQKD \cite{ADFORV18} 
in the most adversarial scenario, where only a minimal set of assumptions (similar to Assumptions \ref{Hyp:DICKA}) is required.
The main point of using the CHSH inequality for cryptographic protocols is due to its self-testing properties, which allows one to derive properties about the devices used during the protocol.
Therefore, in order to prove the security of Device Independent Conference Key Agreement (DICKA)  it is very natural to think of an $N$-partite extension of the CHSH inequality.

In this section we will start by introducing our new $N$-partite Parity-CHSH inequality, which we devise in such a way that it
 closely relates to the CHSH inequality. Then, we presents our DICKA protocol in details and prove
 its security using the connection between our Parity-CHSH inequality and the CHSH inequality.


\subsection*{From CHSH inequality to ``Parity-CHSH'' inequality.}
\label{Sec:P-CHSH}
In this section we present our new Parity-CHSH inequality, that is derived from the CHSH
inequality in such a way that an $N$-partite GHZ state can maximally violate it.

The CHSH inequality \cite{CHSH69} a two-partite inequality that has already proven its usefulness
for device-independent protocols \cite{Acin07,PAB09,VV14,MS14,MS14a,coudron13,KW16,ADFORV18,RPKHW18}. In this section we introduce a slightly different
inequality for $N$ parties. Indeed we use the fact that an $N-$partite GHZ state can be turned
into either $\Phi^+:=\frac{(\ket{00}+\ket{11})(\bra{00}+\bra{11})}{2}$ or $\Phi^-:=\frac{(\ket{00}-\ket{11})(\bra{00}-\bra{11})}{2}$.
by measuring $N-2$ parties in the $X$ basis.
More precisely if the parity of the outcomes of the $N-2$ measurements in the $X$ basis is
$0$ then the state on the remaining $2$ parties is $\Phi^+$,
and if the parity of these outcomes is $1$ then the state on the remaining systems is $\Phi^-$.

The state $\Phi^+$ can be used to maximally violate the CHSH inequality, and $\Phi^-$
can be used to maximally violate an equivalent inequality. Therefore one would expect that
the GHZ state can violate a mixing these two CHSH inequality depending on whether
we create a state $\Phi^+$ or $\Phi^-$.

More precisely, The CHSH inequality can be formulated as a bound on the winning
probability of the following bipartite game.

\begin{Def}[CHSH inequality] \label{Def:CHSH_ine}
 Let Alice and Bob be the two players in this
game called the CHSH game. At the beginning of the game, they are both asked a uniformly random binary question
$x\in\{0,1\}$ and $y\in\{0,1\}$ respectively. They then have to answer bit $a$ and $b$
 respectively. They win the game if and only if \[a+b=xy \text{ mod }2.\]
 No communication is allowed between Alice and Bob during the game. They can, however, agree
 on any strategy before the start of the game. The CHSH inequality states that by using a classical strategy
 -- \emph{i.e.}~modeled with local hidden variables -- Alice and Bob's winning probability must satisfy the following,
\begin{align}P_{\rm win}^{\rm CHSH} \leq \frac{3}{4}.\end{align}
\end{Def}

The state $\Phi^+$ allows to reach the maximum winning probability achievable by quantum mechanics,
\emph{i.e.}~it allows for $P_{\rm win}^{\rm CHSH}=\frac{1}{2} + \frac{1}{2\sqrt{2}} \approx 0.85$.
Similarirly $\Phi^-$ allows to reach the maximum winning probability achievable quantum mechanics($P_{\rm win}\approx 0.85$)
for a game equivalent (up to relabelling) to the CHSH game, in which the winning condition is $a+b=x(y+1) \text{ mod }2.$

Our new ``Parity-CHSH'' inequality extends the CHSH inequality to $N$ parties as follows.

\begin{Def}[Parity-CHSH inequality]\label{Def:new_ineq}
  Let Alice, Bob$_1$, \ldots, Bob$_{N-1}$ be the $N$ players of the following game (the Parity-CHSH game).
Alice and Bob$_1$ are asked uniformly random binary questions $x\in\{0,1\}$ and $y\in\{0,1\}$ respectively. The other
Bobs are each asked a fixed question, \emph{e.g.}~always equal to $1$. The parties all
answer bits $a,b_1,\ldots,b_{N-1}$ respectively. We denote by,
 \[\bar b:=\bigoplus_{2 \leq i\leq N-1} b_i,\]
the parity of the all answers of Bob$_2$,\ldots,Bob$_{N-1}$.
 The players win if and only if
\begin{align}
  a+b_1=x(y+\bar b) \text{ mod }2.
\end{align}
No communication is allowed between Alice and Bob during the game. They can, however, agree on any strategy before the start of the game.
As for the CHSH inequality, classical strategies for the Partity-CHSH game must satisfy,
\begin{align}
    P_{\rm win}^{\rm Parity-CHSH} \leq \frac{3}{4}.
\end{align}
\end{Def}

\begin{Rmk}\label{Rmk:CHSH-PCHSH}
 Note that if we condition on $\bar b=0$, the game is essentially the CHSH game. When conditioned on $\bar b=1$ the
 Parity-CHSH game reduces to a game equivalent to the CHSH up to relabelling the question $y$.
\end{Rmk}

\subsection*{Device Independent Conference Key Agreement}\label{App:DICKA}

We now present a device independent conference key agreement (DICKA) protocol and prove its security in two steps. We first use the
recently developed Entropy Accumulation Theorem \cite{DFR16} to split the overall entropy of Alice's string produced during the
protocol, into a sum of entropy produced on each round of the protocol. Then we use the relation between the Parity-CHSH inequality and the CHSH inequality, to bound the entropy produced
in one round by a function of the violation of the $N$-partite Parity-CHSH inequeality, which generalize the
bounds found for the bipartite case in \cite{PAB09}.

%

\subsection*{The protocol}\label{Sec:Detail_Ptol}

Before we describe our DICKA protocol let us first state the security definitions  for DICKA. We follow the definitions given in \cite{ADFORV18} for DIQKD and generalise it to the multipartite case.
\begin{Def}(Correctness)
   We will call a DICKA protocol $\epsilon_{\rm corr}$-correct for an implementation, if Alice's and Bobs' keys,
  $K_A$, $K_{B_{(1)}},\ldots, K_{B_{(N-1)}}$, are all identical with probability at least $1-\epsilon_{\rm corr}$.
\end{Def}

\begin{Def}(Secrecy)
  We say that a DICKA protocol is $\epsilon_{\rm sec}$-secret for an implementation, if conditioned on not aborting Alice's key $K_A$ is
  $\epsilon_{\rm sec}$-close to a key that Eve is ignorant about. More formally for a key of length $l$, we want
  \[p_{\hat \Omega} \cdot \left\|\rho_{K_A E {|\hat \Omega}} - \frac{\id_A}{2^l}\otimes \rho_{E {|\hat \Omega}}\right\|_{\tr}\leq \epsilon_{\rm sec},\]
  where $\hat \Omega$ is the event of the protocol not aborting, and $p_{\hat \Omega}$ is the probability for $\hat \Omega$.
\end{Def}

Note that if a protocol is $\epsilon_{\rm corr}$-correct and $\epsilon_{\rm sec}$-secret then it is $\epsilon^s$-correct-and-secret for
$\epsilon^s\geq\epsilon_{\rm corr}+\epsilon_{\rm sec}$.

\begin{Def}[Security]\label{Def:security}
  A DICKA protocol is called $(\epsilon^s,\epsilon^c,l)$-secure if:
  \begin{enumerate}
    \item (Soundness) For any implementation of the protocol, either it aborts with probability greater than $1-\epsilon^s$ or it is $\epsilon^s$-correct-and-secret.
    \item (Completeness) There exists a honest implementation of the protocol such that the probability of aborting the protocol is less than $\epsilon^c$, that is
    $1-p_{\hat \Omega} \leq \epsilon^c$.
  \end{enumerate}
\end{Def}

We remark again that Definition \ref{Def:security} was proven to be a criteria for composable security for Quantum Key Distribution in the device dependent scenario \cite{PR14}.
However, for the device independent case it is not known whether such a criteria is enough for composable security. Indeed, Ref.~\cite{noncomposableDIQKD}
suggests that this is not the case if the same devices are used for generation of a subsequent key since this new key can leak information about the first key. Following Ref.~\cite{ADFORV18} we chose to adopt Definition \ref{Def:security}
as the security criteria for DICKA.

We now prove that the DICKA Protocol \ref{PtolA:DICKA}, under the Assumptions \ref{Hyp:DICKA}, satisfies the above definitions of security. For completeness we re-state the protocol here.

\begin{framed}
\begin{Ptol}[More detail version of Protocol \ref{Ptol:DICKA}] \label{PtolA:DICKA}The protocol runs as follows for $N$ parties:
  \begin{enumerate}
    \item For every round $i\in [n]$ do:
      \begin{enumerate}
        \item Alice uses her source to produce and distribute an $N$-partite state, $\rho_{A_i B_{(1\ldots N-1),i}}$, shared among herself and the $N-1$ Bobs.
        \item Alice randomly picks  $T_i$ , s.t. $P(T_i=1)=\mu$, and publicly communicates it to all the Bobs. 
        \item If $T_i=0$ Alice and the Bobs choose $(X_i,Y_{(1...N-1),i})=(0,2,0,...,0)$,
        and if $T_i=1$ Alice chooses $X_i\in_R\{0,1\}$ uniformly at random, Bob$_1$
        chooses $Y_{(1),i}\in_R\{0,1\}$ uniformly at random, and Bob$_2$,\ldots,Bob$_{N-1}$
        choose $(Y_{(2...N-1),i}) =(1,\ldots,1)$.
        \item Alice and the Bobs input the value they chose previously in their respective device and record the output as $ A_i', B'_{(1\ldots N-1),i}$
      \end{enumerate}
    \item They all communicate publicly the list of bases $X_1^n {Y_{(1\ldots N-1)}}_1^n$ they used. \label{Pt_Ptol:comm_round}
    \item \textbf{Error correction:} \label{Error_correction} Alice and the Bobs apply an error correction protocol. Here we chose a protocol based on universal hashing \cite{BS94,RW05}. If the error correction protocol
    aborts for at least one Bob then they abort the protocol. If it does not abort they obtain the raw keys $\tilde K_A, \tilde K_{B_{(1\ldots N-1)}}$. We call $O_A$ the classical information that Alice has sent to the
    Bobs during the error correction protocol. Also the Bobs will send some error correction information but only for the bits produced during the testing rounds ($T_i=1$), for the purpose of parameter estimation.
    We call Alice's guess on Bobs' strings  $G_{(1\ldots N-1)}$, and we denote $O_{(k)}$ the error correction information sent by Bob$_k$.
    \item \textbf{Parameter estimation:} For all the rounds $i$ such that $T_i=1$, Alice uses $A_i'$ and her
    guess on ${B'}_{(1...N-1),i}$ to set $C_i=1$ if they have won the N-partite Parity-CHSH game in the round $i$,
     and she sets $C_i=0$ if they have lost it. Finally she sets $C_i=\bot$ for the rounds $i$ where $T_i=0$.
     She aborts if $\sum_i C_i<\delta \cdot \sum_i T_i$, where $\delta\in]3/4, 1/2+1/2\sqrt{2}[$.
    \item \textbf{Privacy amplification:} Alice and the Bobs apply a privacy amplification protocol (namely the universal hashing described in \cite{RK05}) to
    create final keys $K_A,K_{B_{(1\ldots N-1)}}$. We call $S$ the classical information that Alice sent to the Bobs during the privacy amplification protocol.
  \end{enumerate}
\end{Ptol}
\end{framed}

Note that the above Protocol \ref{PtolA:DICKA} is very similar to the DIQKD protocol given in \cite{ADFORV18}, the difference being that since $N$ parties are present here
we use a shared $N$-partite GHZ state, instead of EPR pairs, and we have to add error corrections.
Indeed we have an error correction protocol
that permits all the parties to get the same raw key. But since we have $N$ parties involved in the protocol, at least one of the parties needs to
know all the other parties' outputs for the testing rounds (when $T_i=1$) in order to estimate, in the
parameter estimation phase, how many times do they succeed in the Parity-CHSH game. For simplicity of the analysis we choose, in Protocol \ref{PtolA:DICKA},
to communicate this information through error correction protocols.

In the ideal scenario (when there is no noise and no interference of Eve) the state $\rho_{A_1^n {B_{(1\ldots N-1)}}_1^n}$ produced corresponds to $n$ copies of the $N$-partite GHZ state,
distributed across the $N$ parties, and Alice and the Bobs measure the following observables:
\begin{enumerate}
  \item Alice's observable for $X_i=0$ is $\sigma_Z$ and for $X_i=1$ it is $\sigma_X.$
  \item Bob$_1$ uses observable $\sigma_Z$ when $Y_{(1),i}=2$, $\frac{\sigma_Z+\sigma_X}{\sqrt{2}}$ when $Y_{(1),i}=0$, and
  $\frac{\sigma_Z-\sigma_X}{\sqrt{2}}$ when $Y_{(1),i}=1$.
  \item For the other Bobs, they have the observable $\sigma_Z$ for $Y_{(k),i}=0$, and for $Y_{(k),i}=1$ they have observable $\sigma_X$.
\end{enumerate}

In the next sections we are going to present the detailed proof of the following main result:

\begin{Thm}\label{Thm:Security}
  Let $\epsilon_{\rm EC},\epsilon'_{\rm EC} \in ]0,1[$ be the two error parameters of the error correction protocol as
  described in the Section \ref{Sec:corr}, $\epsilon_{\rm PA}\in ]0,1[$ be the privacy amplification error probability, $\epsilon_{\rm EA}\in ]0,1[$ be a chosen security parameter for Protocol
  \ref{PtolA:DICKA}, and $\epsilon \in ]0,1[$ be a smoothing parameter.
  Protocol \ref{PtolA:DICKA} is $(\epsilon^s,\epsilon^c,l)$-secure according to Definition \ref{Def:security}, with $\epsilon^s\leq\epsilon_{\rm PA} +2(N-1)\epsilon'_{\rm EC} +2 \epsilon+\epsilon_{\rm EA}$, $\epsilon^c\leq (N-1) (2 \epsilon_{\rm EC}+\epsilon'_{\rm EC}) + \Big(1-\mu \Big( 1- \exp\big[-2 (p_{\rm exp}-\delta)^2\big]\Big)\Big)^n$, and
  \begin{align}\label{eq:key_length}
   \begin{split}
     \hspace{-2mm}l= \!\max_{3/4 \leq \frac{p_{\rm opt}}{\mu} \leq 1/2+1/2\sqrt{2}} \big( (f(\hat q,p_{\rm opt})-\mu)\cdot n - \tilde v \sqrt{n}\big)  &+ 3\log(1-\sqrt{1-(\epsilon/4)^2})-2 \log(\epsilon_{\rm PA}^{-1})\\ &-{\rm leak_{EC}}(O_A)
     - \sum_{k=1}^{N-1} {\rm leak_{EC}}(O_{(k)}),
   \end{split}
   \end{align}
  where $\tilde v= 2\big(\log(13)+ (\hat f'(p_{\rm opt})+1) \big)\sqrt{1-2\log(\epsilon \cdot \epsilon_{\rm EA})}+2 \log(7) \sqrt{-\log(\epsilon_{\rm EA}^2(1-\sqrt{1-(\epsilon/4)^2}))}$, $p_{\rm opt} \in ]\mu 3/4,\mu (1/2+1/2\sqrt{2})[$
  is a parameter to be optimized: more precisely $p_{\rm opt}$ is the unique point were the tangent function $f(\,\cdot\,,p_{\rm opt})$ to the function $\hat f(\cdot)$ (see Lemma~\ref{Lemma:f}) is such that $f(p_{\rm opt},p_{\rm opt})= \hat f(p_{\rm opt})$ (by convexity of $\hat f$ we have $\forall x \in[0,1]\ f(x,p_{\rm opt})\leq \hat f(x)$).
  Finally $p_{\rm exp}$ is the expected winning probability to win a single round of the Parity-CHSH game for a honest implementation, $\delta \in ]3/4,1/2+1/2\sqrt{2}[$ is the threshold defined in Protocol \ref{PtolA:DICKA}, and $\hat q$ is the vector $(\mu\delta,\mu-\mu \delta,1-\mu)^t$.
\end{Thm}


\subsection*{Correctness}\label{Sec:corr}
The correctness of Protocol \ref{PtolA:DICKA} comes from the first part of the error  correction protocol used by the parties, where Alice sends information to the Bobs so that they generate the raw keys $\tilde K_A, \tilde K_{B_{(1\ldots N-1)}}$. We want here an error correction protocol
that uses only communication from Alice to the Bobs and that minimizes the amount of communication needed. Therefore we are going to use 
an error correction protocol as the one
described in \cite{BS94,RW05}. The idea of this error correction code is that Alice chooses a hash function and sends to the Bobs the chosen function and the hashed value of her bits. We denote this communication $O_A$. Then
each Bob$_k$ will individually use $O_A$ and his own prior knowledge ${B_{(k)}}_1^n {X_A}_1^n {Y_{{(1\ldots N-1)}}}_1^n T_1^n$ to guess  Alice's string. Each of the Bobs can fail to produce a guess, so if one of
them fails the protocol aborts. In an honest implementation of the protocol, the probability that one particular Bob, say Bob$_k$ ($k\in[N-1]$), aborts is upper bounded by $\epsilon_{\rm EC}$.
Therefore the probability that at least one of them aborts in an honest implementation is at most $(N-1) \epsilon_{\rm EC}$. If for $k \in [N-1]$ Bob$_k$
does not abort we then have that $P(\tilde K_A \neq  \tilde K_{B_{(k)}})\leq \epsilon'_{\rm EC}$. Therefore if none of the Bobs aborts we have that,
\begin{align*}
  P\big(\tilde K_A = \tilde K_{B_{(1)}}=\ldots=\tilde K_{B_{(N-1)}}\big)&=1-P\big(\tilde K_A \neq \tilde K_{B_{(1)}}\ {\rm OR} \ldots {\rm OR}\ \tilde K_A \neq \tilde K_{B_{(N-1)}}\big) \\
  &\geq 1- (N-1)\epsilon'_{\rm EC} \geq1-\epsilon_{\rm corr},
\end{align*}
where we take $\epsilon_{\rm corr} \geq (N-1)\epsilon'_{\rm EC}$, which proves the following lemma:

\begin{Lmm}\label{Lm:correctness}
  The Protocol \ref{PtolA:DICKA} is $\epsilon_{\rm corr}$-correct, for any $\epsilon_{\rm corr}\geq (N-1)\epsilon'_{\rm EC}$, where $\epsilon'_{\rm EC}$ is such that if $\forall k \in[N-1]$ Bob$_k$ does not abort the error correction protocol then
  $P(\tilde K_A \neq  \tilde K_{B_{(k)}})\leq \epsilon'_{\rm EC}$ .
\end{Lmm}

\subsection*{Completeness}

We call an \emph{honest implementation} of the protocol, an implementation where the measurement devices used act in the same way in all the rounds of the protocol,
 the state used for the $n$ rounds is of the form $\rho_{A B_{(1\ldots N-1)}}^{\otimes n}$ (the measurements and the state are then said to be identically and independently distributed (i.i.d.)),
 and such that for one single round, the probability of winning
 the $N$ partite Parity-CHSH game is $p_{\rm exp} \allowbreak \in \big] 3/4, 1/2+1/2\sqrt{2}\big]$.

  \begin{Lmm}\label{Lmm:completness}
    For any parameter $\delta \in \left] 3/4, 1/2+1/2\sqrt{2}\right[$, Protocol \ref{PtolA:DICKA} is $\epsilon^c$-complete, for
     \begin{align}
     \epsilon^c\leq (N-1) (2 \epsilon_{\rm EC}+\epsilon'_{\rm EC}) + \Big(1-\mu \Big( 1- \exp\big[-2 (p_{\rm exp}-\delta)^2\big]\Big)\Big)^n ,
   \end{align}
     where $p_{\rm exp} > \delta$, $\delta$ is a threshold.
  \end{Lmm}
  \begin{proof}
    Protocol \ref{PtolA:DICKA} can abort at two moments: it can abort during the error correction or during the parameter estimation.
    For the error correction step, the protocol aborts if one of the Bobs aborts while trying to guess Alice's string, or if Alice aborts while guessing
    Bobs' testing bits. We are assuming that the Bobs use the same error correction protocol in order to send information about their outputs in the test rounds so that Alice can make her guess.
     Therefore the overall probability of aborting during the error correction protocol is then bounded by $2(N-1)\epsilon_{\rm EC}$
    for an honest implementation.
    The probability of aborting during the parameter estimation part (conditioned on not aborting the error correction step) is given by:
    \begin{align}
      \begin{split}
      \hspace{-3mm}P_{\rm PE}({\rm abort})&=P(G_{(1\ldots N-1)}\text{ is correct}) P\Big(\sum_{i} C_i < \delta \cdot \sum_i T_i\Big| G_{(1\ldots N-1)}\text{ is correct}\Big) \\
      &\hspace{1.7cm} +P(\exists k: G_{(k)}\text{is wrong}) P\Big(\sum_{i} C_i < \delta \cdot \sum_i T_i\Big| \exists k: G_{(k)}\text{is wrong}\Big)
    \end{split}
    \end{align}
    where $G_{(k)}$ is Alice's guess for Bob$_k$'s testing rounds bits. It is said to be correct when the string $G_{(k)}={B'}_{(k),I}$ for
    $I:=\{i\in [n]:\, T_i=1\}$. By bounding $P(G_{(1\ldots N-1)}\text{ is correct})$ by $1$, $P(\exists k: G_{(k)}\text{is wrong})$ by $(N-1) \epsilon'_{\rm EC}$, and
       $P\Big(\sum_{i} C_i < \delta \cdot \sum_i T_i\Big| \exists k: G_{(k)}\text{is wrong}\Big)$ by $1$, we get
    \begin{align}
      P_{\rm PE}({\rm abort}) \leq \sum_{j=0}^n P\Big(\sum_i T_i = j\Big)\cdot P\Big(\sum_{i} C_i < \delta \cdot j \Big| \sum_i T_i = j\ \&\ \forall k \tilde K_A= \tilde K_{B_{(k)}}\Big) + (N-1)\epsilon_{\rm EC}'. \label{eq:bound_P_PE}
    \end{align}

    Let us consider an honest implementation such that $p_{\rm exp} > \delta$, we can then rewrite \eqref{eq:bound_P_PE} as,
    \begin{align}
      \begin{split}
      P_{\rm PE}&({\rm abort}) \label{eq:bound_p_abort}\\
      &\leq \sum_{j=0}^n P\Big(\sum_i T_i = j\Big)\cdot P\Big(\sum_{i} C_i < \big(p_{\rm exp}-(p_{\rm exp} -\delta)\big) \cdot j \Big| \sum_i T_i = j\ \&\
      G_{(1\ldots N-1)}\text{ is correct}\Big)\\ &\hspace{10cm} +(N-1)\epsilon'_{\rm EC}.
    \end{split}
    \end{align}
    Note that the expectation value $\mathbb{E}(C_i)= p_{\rm exp}$ and because an honest implementation is i.i.d. we can use Hoeffding
    inequalities to bound $P\Big(\sum_{i} C_i < \big(p_{\rm exp}-(p_{\rm exp} -\delta)\big) \cdot j \Big| \sum_i T_i = j\ \&\ G_{(1\ldots N-1)}\text{ is correct}\Big)< \exp(-2 (p_{\rm exp}-\delta)^2 j)$. Moreover the the i.i.d.~random variables
    $T_i$ follow a Bernoulli distribution with $P(T_i=1)=\mu$. Pluging all of this into eq.~\eqref{eq:bound_p_abort} gives us,
    \begin{align}
      P_{\rm PE}({\rm abort})&\leq \sum_{j=0}^n \binom{n}{j} (1-\mu)^{n-j} \mu^j \times \exp(-2 (p_{\rm exp}-\delta)^2 j) +(N-1)\epsilon'_{\rm EC}\\ &=
      \sum_{j=0}^n \binom{n}{j} (1-\mu)^{n-j} \big(\mu \times \exp(-2 (p_{\rm exp}-\delta)^2 )\big)^j +(N-1)\epsilon'_{\rm EC}\\
      &= \Big(1-\mu \Big( 1- \exp\big[-2 (p_{\rm exp}-\delta)^2\big]\Big)\Big)^n +(N-1)\epsilon'_{\rm EC},
    \end{align}
  where the last equality comes from the binomial theorem.
  \end{proof}

\subsection*{Soundness}

In order to complete the security proof of Protocol \ref{PtolA:DICKA}, it
remains to prove secrecy. Let $\hat \Omega'$ be the event that  Protocol~\ref{PtolA:DICKA} does not abort and that the error correction step is successful.
The Leftover Hashing Lemma \cite[Corollary 5.6.1]{RennerThesis}  states that the secrecy of the final key, after a privacy amplification protocol using a family of two-universal hashing functions, depends on the amount of smooth min-entropy of the state before privacy amplification conditioned on the event $\hat \Omega'$.

\begin{Thm}[Leftover Hashing Lemma \cite{RennerThesis}]\label{thm:leftover}
Let $\mathcal{F}$ be a family of two-universal hashing functions from $\{0,1\}^n \rightarrow \{0,1\}^l$, such that $F(A_1^n)= K_A$ for $F \in \mathcal{F}$, then it holds that
\begin{align}
 \left\|\rho_{K_A E {|\hat \Omega'}} - \frac{\id_A}{2^l}\otimes \rho_{E {|\hat \Omega'}}\right\|_{\tr} \leq 2\epsilon+2^{-\frac{1}{2}(H_{\min}^\epsilon({A}_1^n|E)_{\rho_{|\hat \Omega'}}-l)}.
 \end{align}
\end{Thm}

According to Theorem~\ref{thm:leftover}, in order to  prove the secrecy of Protocol \ref{PtolA:DICKA} we need to lower bound the
smooth min-entropy $H_{\min}^\epsilon({A'}_1^n|X_1^n {Y_{(1\ldots N-1)}}_1^n T_1^n O O_{(1\ldots N-1)} E)_{\rho_{|\hat \Omega'}}$.
The proof goes in the following steps: In Lemma \ref{Lemma:EAT_ECsucc}, we introduce an error correction map and bound the entropy $H_{\min}^{\epsilon} ({A'}_1^n | X_1^n {Y_{(1\ldots N-1)}}_1^n  T_1^n E)$
for the state after the action of the error correction map, conditioned on the event that a particular violation is observed and the error correction protocol is successful.
 In Lemma \ref{Lmm:bound_entropy}, we relate the state generated by Protocol~\ref{PtolA:DICKA}  conditioned on the event that the error correction protocols were successful to the state artificially introduced in Lemma \ref{Lemma:EAT_ECsucc},
and we estimate $H_{\min}^{\epsilon} ({A'}_1^n | X_1^n {Y_{(1\ldots N-1)}}_1^n  T_1^n O O_{(1\ldots N-1)} E)$, taking into account the information leaked during the error correction protocol.
 Finally, in Lemma \ref{Lmm:soundnes},  we combine the previous results proving the soundness of Protocol~\ref{PtolA:DICKA}.

To bound the smooth min-entropy we will use the EAT. Indeed, before the error correction part, Protocol~\ref{PtolA:DICKA} can be described by a
composition of EAT channels that we will call $\mathcal{M}_1,\ldots,\mathcal{M}_n$ (see Fig.~\ref{fig:maps}).

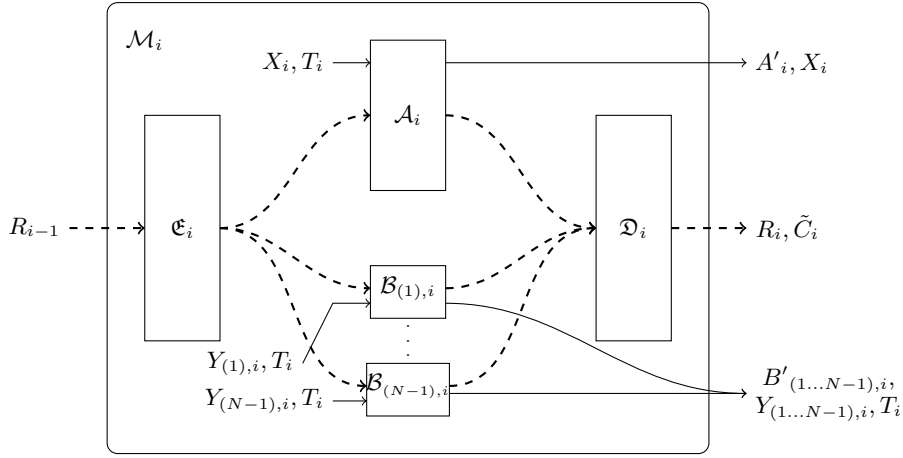
\begin{figure}[h]
  \center
  \begin{tikzpicture}[scale=1]
    \draw[rounded corners] (-4,-3) rectangle (4,3);
    \node[] at (-3.5,2.5) {$\mathcal{M}_i$};
    \draw[] (-3.5,-1.5) rectangle (-2.5,1.5);
    \node[] at (-3,0) {$\mathfrak{E}_i$};
    \draw[] (3.5,-1.5) rectangle (2.5,1.5);
    \node[] at (3,0) {$\mathfrak{D}_i$};
    \draw[] (-0.5,0.5) rectangle (0.5,2.5);
    \node[] at (0,1.5){$\mathcal{A}_i$};
    \draw[] (-0.5,-0.5) rectangle (0.5,-1.2);
    \node[] at (0,-0.8){ $\mathcal{B}_{(1),i}$};
    \node[thick] at(0,-1.5){$\substack{\cdot\\ \cdot\\ \cdot}$};
    \draw[] (-0.55,-1.8) rectangle (0.55,-2.5);
    \node[] at (0,-2.1){ $\mathcal{B}${\tiny$_{(N-1),i}$}};
    \draw[->,thick, dashed] (-4.5,0)node[left] {$R_{i-1}$} to (-3.5,0);
    \draw[->,out=0, in=180,thick, dashed] (-2.5,0) to (-0.5,1.5);
    \draw[->,out=0, in=180,thick, dashed] (-2.5,0) to (-0.5,-0.8);
    \draw[->,out=0, in=180,thick, dashed] (-2.5,0) to (-0.55,-2.1);
    \draw[->,out=0, in=180,thick, dashed] (0.5,1.5) to (2.5,0);
    \draw[->,out=0, in=180,thick, dashed] (0.55,-2.1) to (2.5,0);
    \draw[->,out=0, in=180,thick, dashed] (0.5,-0.8) to (2.5,0);
    \draw[->,thick, dashed] (3.5,0) to (4.5,0) node[right]{$R_i,\tilde C_i$};
    \draw[->] (-1,2.2)node[left]{$X_i,T_i$} to (-0.5,2.2);
    \draw[->] (-1.4,-1.8)node[left]{$Y_{(1),i}, T_i$} -- (-1,-1)-- (-0.5,-1);
    \draw[->,out=0,in=180] (0.5,-1) to (4.5,-2.2);
    \draw[->] (-1,-2.3)node[left]{$Y_{(N-1),i}, T_i$} to (-0.55,-2.3);
    \draw[->](0.5,2.2) -- (4.5,2.2)node[right]{${A'}_i,X_i$};
    \draw[->] (0.55,-2.2) to (4.5,-2.2)node[right,align=center]{${B'}_{(1\ldots N-1),i}$,\\$Y_{(1\ldots N-1),i},T_i$};
  \end{tikzpicture}
  \caption{Description of the map $\mathcal{M}_i$. This map describes the round $i$ of the first step
  of the Protocol~\ref{PtolA:DICKA}. $T_i$ is chosen at random such that
    $P(T_i=1)=\mu$. $X_i \in
    \{0,1\}$ represents the ``basis'' in which Alice's device, represented by the CPTP map
    $\mathcal{A}_i$, measures its input to get the output $A'_i \in
    \{0,1\}$. $X_i=0$ when $T_i=0$ and $X_i \in_R \{0,1\}$ otherwise.
     $ Y_{(k),i}\in \{0,1,2\}$ represents the ``basis'' in which
    Bob$_k$'s device, represented by the CPTP map $\mathcal{B}_{(k),i}$, measures its input to get the
    output $B'_{(k),i} \in \{0,1\}$. If $T_i=0$ we have
    $Y_{(k),i}=2$, else we have $Y_{(k),i} \in_R \{0,1\}$. If $T_i=0$
    then $\tilde C_i=\bot$, else $\tilde C_i=w_{\rm Parity-CHSH}(A'_i,B'_{(1\ldots N-1),i},X_i,Y_{(1\ldots N-1),i})$.}
  \label{fig:maps}
\end{figure}

In order to apply EAT we need to find a min-tradeoff function for the maps $\mathcal{M}_i$ defined by the Figure \ref{fig:maps}. \textit{I.e.}, we need to find a function $f$ such that
\begin{align}
  f(q) \leq \inf_{\sigma \in \Sigma_i(q)} H(A'_i \tilde C_i| X_i Y_{(1\ldots N-1),i} T_i R)_{\sigma},
\end{align}
for
\[ \Sigma_i(q):= \{\sigma_{\tilde C_i A'_i B'_{(1\ldots N-1),i} X_i Y_{(1\ldots N-1),i} T_i R_i R}= (\mathcal{M}_i\otimes \id_R)(\sigma_{R_{i-1} R}): \sigma_{R_{i-1} R} \in \mathcal{S}(\mathcal{H}_{R_{i-1}R})\ \&\ \sigma_{\tilde C_i}=q\},\]
where $\Sigma_i(q)$ is the set of states that can be generated by the action of the channel $\mathcal{M}_i\otimes \id_R$ on an arbitrary state and such that the classical register $\tilde{C}_i$ has distribution $q$.

\begin{Lmm}\label{Lemma:f}
  The real function defined as,
  \begin{align}
\hat f(x):=\left(1-\frac{\mu}{2}\right)\left(1-h\left(\frac{1}{2} + \frac{1}{2}\sqrt{(4x/\mu-2)^2-1}\right) \right)
  \end{align}
  is a min-tradeoff function for the EAT channels $\mathcal{M}_i$ defined by the Figure \ref{fig:maps}.
Here $\mu$ is the testing probability of the Protocol \ref{PtolA:DICKA}, and
  $h(x)$ is the binary entropy: $h(x)=-x\log(x)-(1-x)\log(1-x)$.\\
  We define the affine function $f(\ \cdot\ ,p_{\rm opt})$ over the probability distribution $\mathds{P}(\{1,0,\bot\})$ as, $\forall\ q=\big(q(1),q(0),q(\bot)\big)^t\in \mathds{P}(\{1,0,\bot\}),$
  \begin{align}
    &f(q,p_{\rm opt}):=\hat f'(p_{\rm opt}) q(1) + \hat f(p_{\rm opt})-\hat f'(p_{\rm opt}) p_{\rm opt},
  \end{align}
  where $p_{\rm opt} \in ]\mu 3/4, \mu (1/2+1/2\sqrt{2})[$.
\end{Lmm}

In order to make the argument
more rigorous and general, here $f(\cdot,p_{\rm opt})$ is a function that takes as input the vector of frequencies $q=\big(q(1),q(0),q(\bot)\big)^t$.

\begin{proof}
Let us take a state $\sigma_{\tilde C_i A'_i B'_{(1\ldots N-1),i} X_i Y_{(1\ldots N-1),i} T_i R_i R} \in \Sigma_i(q)$. Then we define the state
\begin{align}
  \sigma'_{\tilde C_i A''_i B''_{(1),i}B'_{(2\ldots N-1),i} X_i Y_{(1\ldots N-1),i} T_i F_i R_i R}
\end{align}
to be the state we obtain from $\sigma_{\tilde C_i A'_i B'_{(1\ldots N-1),i} X_i Y_{(1\ldots N-1),i} T_i R_i R}$ by replacing $A'_i$ by
$A''_i:= A'_i \oplus F_i$ and $B'_{(1),i}$ by $B''_{(1),i}:=B'_{(1),i} \oplus F_i$ where $F_i$ is a bit that is chosen uniformly at random.
None of the other registers are changed, in particular, note that we still have that $\sigma'_{\tilde C_i}=q$, where the value of $\tilde{C}_i$ can be determined by the registers $A''_i$, $B''_{(1),i}$, and $B'_{(2\ldots N-1),i}$. Moreover, since $F_i$ is completely independent
of the other variables and given the definition of $A''_i$, it is easy to check that,
\begin{align}
   H(A'_i \tilde C_i| X_i Y_{(1\ldots N-1),i} T_i R)_{\sigma} = H(A''_i \tilde C_i| F_i X_i Y_{(1\ldots N-1),i} T_i R)_{\sigma'}.
\end{align}
This entropy can be lower bounded as follows:
\begin{align}
  H(A''_i \tilde C_i| F_i X_i Y_{(1\ldots N-1),i} T_i R)_{\sigma'} \geq H(A''_i \tilde C_i| F_i X_i Y_{(1\ldots N-1),i} T_i R, \bar b)_{\sigma'},
\end{align}
where $\bar b$ denotes the register containing the parity of ${B'}_{(2\ldots N-1),i}$. The right-hand side of the above inequality can be
expended as,
\begin{align}
  \label{eq:expand_H}H(A''_i \tilde C_i| F_i X_i Y_{(1\ldots N-1),i} T_i R, \bar b)_{\sigma'} = p_{\bar b=0} &H(A''_i \tilde C_i| F_i X_i Y_{(1\ldots N-1),i} T_i R, \bar b=0)_{\sigma'}\\
  \notag&+ p_{\bar b=1} H(A''_i \tilde C_i| F_i X_i Y_{(1\ldots N-1),i} T_i R, \bar b=1)_{\sigma'}
\end{align}
We will now detail the derivation of a lower bound for $H(A''_i \tilde C_i| F_i X_i Y_{(1\ldots N-1),i} T_i R, \bar b=0)$. The lower bound
on $H(A''_i \tilde C_i| F_i X_i Y_{(1\ldots N-1),i} T_i R, \bar b=1)$ follows the exacte same steps.

Using the chain rule,
\begin{align}
    H(A''_i \tilde C_i| F_i X_i Y_{(1\ldots N-1),i} T_i R, \bar b=0)_{\sigma'} \geq H(A''_i | F_i X_i Y_{(1\ldots N-1),i} T_i R, \bar b=0)_{\sigma'},
\end{align}
and since $P(X_i=0)=1-\frac{\mu}{2}$,
\begin{align}
  H(A''_i | F_i X_i Y_{(1\ldots N-1),i} T_i R, \bar b=0)_{\sigma'} \geq \left(1-\frac{\mu}{2}\right)\cdot H(A''_i|F_i Y_{(1\ldots N-1),i} T_i R,X_i=0, \bar b=0)_{\sigma'}.
\end{align}
Given that for $X_i=0$ Alice's measurement is independent of $Y_{(1\ldots N-1),i}$ and $T_i$ we have
\begin{align}
  H(A''_i|F_i Y_{(1\ldots N-1),i} T_i R,X_i=0, \bar b=0)_{\sigma'} = H(A''_i|F_i R,X_i=0, \bar b=0)_{\sigma'}.
\end{align}
Using the definition of the conditional Von Neumann entropy we can write:
  \begin{align}
    H(A''_i|F_i R,X_i=0, \bar b=0)_{\sigma'} &= H(A''_i F_i R |X_i=0, \bar b=0)_{\sigma'} -H(F_i R |X_i=0, \bar b=0)_{\sigma'}\\
    &\hspace{-1.3cm}= H(A''_i| X_i=0, \bar b=0)_{\sigma'}+ \underbrace{H( F_i R |X_i=0, \bar b=0)_{\sigma'}-H(F_i R |X_i=0, \bar b=0)_{\sigma'}}_{=-\chi(A''_i:F_i R|X=0, \bar b=0)_{\sigma'}}\\
    &= 1-\chi(A''_i:F_i R|X_i=0, \bar b=0)_{\sigma'}
  \end{align}
  where $\chi(A''_i:F_i R|X_i=0, \bar b=0)$ is the Holevo quantity, and the last equality comes from the definition of $A''_i$ being a uniform variable (for any value of $X_i$ and $\bar b$).

In the following we will use $p_w$ for $P_{\rm win}^{\rm Parity-CHSH}$. From the definition of the Parity-CHSH inequality (see Def.~\ref{Def:new_ineq}),
one can noticed that conditioned on $\bar b=0$, the Parity-CHSH game reduces the the usual CHSH game, and conditioned on $\bar b=1$ it reduces to a game
that is equivalent to CHSH up to flipping input $y$. Therefore we can write $p_w=p_{\bar b=0} p_{w|\bar b=0} + p_{\bar b=1} p_{w|\bar b=1}$,
where $p_{w|\bar b=0}$ can be viwed as the winning probability of a CHSH game, and $p_{w|\bar b=1}$ as the winning probability of the CHSH game with flipped
input $y$.
Moreover, for any state leading to a CHSH violation of $P_{\rm win}^{\rm CHSH} = p_{w|\bar b=0} \in [3/4,1/2+1/2\sqrt{2}]$,
Ref.~\cite[Section 2.3]{PAB09} gives a tight upper bound on $\chi(A_i:F_i R|X_i=0, \bar b=0)$:
\begin{align}
  \chi(A_i'':RF_i|X_i=0, \bar b=0) \leq h\left( \frac{1}{2}+ \frac{1}{2}\sqrt{(4p_{w|\bar b=0}-2)^2-1}\right), \label{eq:bound_chi_CHSH}
\end{align}
where $h(x)=-x\log(x)-(1-x)\log(1-x)$.
This leads to,
\begin{align}
  H(A''_i \tilde C_i| F_i X_i Y_{(1\ldots N-1),i} T_i R, \bar b=0)_{\sigma'} \geq 1- h\left( \frac{1}{2}+ \frac{1}{2}\sqrt{(4p_{w|\bar b=0}-2)^2-1}\right).
\end{align}
Similarily we can bound,
\begin{align}
  H(A''_i \tilde C_i| F_i X_i Y_{(1\ldots N-1),i} T_i R, \bar b=1)_{\sigma'} \geq 1- h\left( \frac{1}{2}+ \frac{1}{2}\sqrt{(4p_{w|\bar b=1}-2)^2-1}\right).
\end{align}
Using these two above inequalities in eq.~\eqref{eq:expand_H} we get,
\begin{align}
  H(A''_i \tilde C_i| F_i X_i Y_{(1\ldots N-1),i} T_i R, \bar b)_{\sigma'} &\geq p_{\bar b=0} \left(1- h\left( \frac{1}{2}+ \frac{1}{2}\sqrt{(4p_{w|\bar b=0}-2)^2-1}\right)\right)\\
  \notag &\hspace{1cm}+ p_{\bar b=1}  \left(1- h\left( \frac{1}{2}+ \frac{1}{2}\sqrt{(4p_{w|\bar b=1}-2)^2-1}\right)\right)\\
  &\geq \left(1- h\left( \frac{1}{2}+ \frac{1}{2}\sqrt{(4(p_{\bar b=0} p_{w|\bar b=0} + p_{\bar b=1} p_{w|\bar b=1})-2)^2-1}\right)\right)\\
  &=\left(1- h\left( \frac{1}{2}+ \frac{1}{2}\sqrt{(4 p_w-2)^2-1}\right)\right),
\end{align}
where the last inequality holds by convexity.

Note that $p_w$ can be expressed in terms of the probability distribution $q=(q(1),q(0),q(\bot))^t$ (where $^t$ is the transpose) as
$p_w=\frac{q(1)}{1-q(\bot)}$. And because in our case the definition of the maps $\mathcal{M}_i$ implies $1-q(1)=\mu$ we have $p_w=\frac{q(1)}{\mu}$. Therefore the function
\begin{align}
  \bar f(q)=\hat f(q(1))=\left(1-\frac{\mu}{2}\right)\cdot \left(1-h\left( \frac{1}{2}+ \frac{1}{2}\sqrt{\big(4 \cdot q(1)/\mu -2\big)^2-1}\right) \right),
\end{align}
is a min-tradeoff function, and $\hat f$ is a differentiable convex increasing function of one variable.
To find an affine min-tradeoff function $f$ we take a tangent to $\hat f$ for some value $p_{\rm opt}(n,\delta) \in \left]\mu \cdot 3/4,\mu \cdot (1/2+1/2\sqrt{2})\right[$ to be chosen, where $\mu$ and $\delta$ are
defined in the Protocol \ref{PtolA:DICKA}, which gives us,
\begin{align}
  f(q,p_{\rm opt}):= \hat f'(p_{\rm opt}) q(1) + \hat f(p_{\rm opt})-\hat f'(p_{\rm opt}) p_{\rm opt}.
\end{align}
\end{proof}

In the following Lemma we show that the state $\tilde \rho$ created by applying a sequence of $n$ CPTP maps of
the form described by Fig.~\ref{fig:maps} on some initial state, (when conditioned on
 the event of having (statistically) high enough Bell violation),
possesses a linear amount of entropy.

\begin{Lmm}\label{Lemma:EAT_ECsucc}
   Let $\mathcal{M}_{\rm EC}$
  be the CPTP map $ {A'}_1^n {{B'}_{(1\ldots N-1)}}_1^n \mapsto {A'}_1^n {{B'}_{(1\ldots N-1)}}_1^n K_{B_{(1\ldots N-1)}} G_{(1\ldots N-1)}$ that models the error correction protocols, applied during Step \ref{Error_correction} of
  Protocol \ref{PtolA:DICKA},  which produce the raw keys $K_{B_{(1\ldots N-1)}}$ and the guess $G_{(1\ldots N-1)}$. For $i\in [n]$ let $\mathcal{M}_i$ be the CPTP map from $R_{i-1}$ to ${A'}_i {{B'}_{(1\ldots N-1)}}_i \tilde C_i X_i Y_{(1\ldots N-1),i} T_i R_i$
  defined in the Fig.~\ref{fig:maps}. Let $\Omega$ be the event \{$\sum_j \tilde C_j \geq \delta \cdot \sum_j T_j$ for $\delta \in \left] 3/4, 1/2+1/2\sqrt{2}\right[$
  \textbf{and} all the error correction protocols were successful, meaning that $\forall k,\ {A'}_1^n=K_{B_{(k)}}$ and Alice guess $G_{(1\ldots N-1)}$ is correct\}.
  We define the state,
  \begin{align}
    \tilde \rho_{{A'}_1^n \tilde C_1^n {{B'}_{(1\ldots N-1)}}_1^n X_1^n {Y_{(1\ldots N-1)}}_1^n T_1^n E} := (\tr_{R_n} \circ \mathcal{M}_n \circ \ldots \circ \mathcal{M}_1 \otimes \id_E)(\rho_{R_0 E}),
  \end{align}
  where $R_0= A_1^n {B_{(1\ldots N-1)}}_1^n$, and $\rho_{R_0 E}$ is the state shared between Alice, the Bobs, and Eve (produced by Alice's source) across the $n$ rounds of the Protocol \ref{PtolA:DICKA} before they apply any measurement.
  Then we have for any $\epsilon \in ]0,1[$,
  \begin{align}
    H_{\min}^\epsilon\big({A'}_1^n | X_1^n &{Y_{(1\ldots N-1)}}_1^n  T_1^n E\big)_{\mathcal{M}_{\rm EC}(\tilde \rho)_{|\Omega}}\geq \big(f(\hat q,p_{\rm opt})-\mu\big)\cdot n - \tilde v \sqrt{n} + 3\log(1-\sqrt{1-(\epsilon/4)^2}), \label{eq:entrop_ineq}
  \end{align}
  where $\tilde v= 2\big(\log(13)+ (\hat f'(p_{\rm opt})+1) \big)\sqrt{1-2\log(\epsilon \cdot p_{\Omega})}+2 \log(7) \sqrt{-\log(p_\Omega^2(1-\sqrt{1-(\epsilon/4)^2}))}$, and
  $\hat q = \big(\delta \mu,\mu-\delta \mu, 1-\mu \big)^t\in \mathds{P}(\{1,0,\bot\})$.
\end{Lmm}
\begin{proof}
  \noindent Note that $\tilde \rho_{|\Omega}:= \tr_{K_{B_{(1\ldots N-1)}} G_{(1\ldots N-1)}}\big(\mathcal{M}_{\rm EC}(\tilde \rho)_{|\Omega}\big)$, therefore
  \linebreak $H_{\min}^\epsilon\big({A'}_1^n|X_1^n {Y_{(1\ldots N-1)}}_1^n  T_1^n E\big)_{\mathcal{M}_{\rm EC}(\tilde \rho)_{|\Omega}}= H_{\min}^\epsilon\big({A'}_1^n|X_1^n {Y_{(1\ldots N-1)}}_1^n  T_1^n E\big)_{\tilde \rho_{|\Omega}}$.

  The maps $\mathcal{M}_1,\ldots, \mathcal{M}_n$ are EAT channels with the following Markov conditions,
  \begin{align}
    \forall i \in [n], {A'}_1^{i-1} \tilde C_1^{i-1}\leftrightarrow X_1^{i-1} {Y_{(1\ldots N-1)}}_1^{i-1} T_1^{i-1} E \leftrightarrow X_1^{i} {Y_{(1\ldots N-1)}}_1^{i} T_1^{i}.
  \end{align}

  Indeed for any round $i \in [n]$ the variables $X_i Y_{(1\ldots N-1),i} T_i$ are chosen independently of any other round $j \neq i$.
  We have proven  that the function $f(\ \cdot\ ,p_{\rm opt})$  is a min-tradeoff
  function for the maps $\mathcal{M}_1,\ldots,\mathcal{M}_n$. We can therefore use the EAT to bound \linebreak  $H_{\min}^\epsilon\big({A'}_1^n \tilde C_1^n| X_1^n {Y_{(1\ldots N-1)}}_1^n  T_1^n E\big)_{\tilde \rho_{|\Omega}}$:
   \begin{align}
    H_{\min}^\epsilon\big({A'}_1^n \tilde C_1^n| X_1^n {Y_{(1\ldots N-1)}}_1^n  T_1^n E\big)_{\tilde \rho_{|\Omega}} \geq n f(\hat q, p_{\rm opt}) - c \sqrt{n}, \label{eq:EAT_bound}
  \end{align}
  where $\hat q=(\mu \delta, \mu-\mu\delta,1-\mu)$, $c= 2\big(\log(13)+ \lceil \hat f'(p_{\rm opt})\rceil \big)\sqrt{1-2\log(\epsilon \cdot p_{\Omega})}$, and $p_{\Omega}$ is the probability of the event $\Omega$.
  This is true because $f(q, ,p_{\rm opt})$ is an increasing function of $q(1)$, so for any event that implies $\sum_j \tilde C_j \geq \delta \cdot \sum_j T_j$ we have that $f(\freq{\tilde C_1^n},p_{\rm opt})\geq f(\hat q, p_{\rm opt})$,
  in particular $\Omega \Rightarrow f(\freq{\tilde C_1^n},p_{\rm opt})\geq f(\hat q, p_{\rm opt})$. Note that because $\forall x\in \mathbb{R},\ \lceil x \rceil \leq x+1$ we can upper bound
  $\lceil \hat f'(p_{\rm opt})\rceil$ by $\hat f'(p_{\rm opt})+1$ and then take $c= 2\big(\log(13)+ (\hat f'(p_{\rm opt})+1) \big)\sqrt{1-2\log(\epsilon \cdot p_{\Omega})}$.

  Using \cite[eq.~(6.57)]{T15} we can relate $H_{\min}^\epsilon\big({A'}_1^n \tilde C_1^n| X_1^n {Y_{(1\ldots N-1)}}_1^n  T_1^n E\big)_{\tilde \rho_{|\Omega}}$ to \linebreak
  $H_{\min}^\epsilon\big({A'}_1^n | X_1^n {Y_{(1\ldots N-1)}}_1^n  T_1^n E\big)_{\tilde \rho_{|\Omega}}$:
  \begin{align}
    \begin{split}
    \hspace{-5mm}H_{\min}^\epsilon\big({A'}_1^n | X_1^n &{Y_{(1\ldots N-1)}}_1^n  T_1^n E\big)_{\tilde \rho_{|\Omega}}  \\
    \geq & H_{\min}^{\frac{\epsilon}{4}}\big({A'}_1^n \tilde C_1^n| X_1^n {Y_{(1\ldots N-1)}}_1^n  T_1^n E\big)_{\tilde \rho_{|\Omega}}
    -H_{\max}^{\frac{\epsilon}{4}}\big( \tilde C_1^n| X_1^n {Y_{(1\ldots N-1)}}_1^n  T_1^n E\big)_{\tilde \rho_{|\Omega}} \\
    &\hspace{6.2cm}+ 3\log(1-\sqrt{1-(\epsilon/4)^2}). \label{eq:bound1}
  \end{split}
  \end{align}

  We now need to upper bound $H_{\max}^{\frac{\epsilon}{4}}\big( \tilde C_1^n| X_1^n {Y_{(1\ldots N-1)}}_1^n  T_1^n E\big)_{\tilde \rho_{|\Omega}}$. First we note that,
  \[ H_{\max}^{\frac{\epsilon}{4}}\big( \tilde C_1^n| X_1^n {Y_{(1\ldots N-1)}}_1^n  T_1^n E\big)_{\tilde \rho_{|\Omega}} \leq H_{\max}^{\frac{\epsilon}{4}}\big( \tilde C_1^n|  T_1^n E\big)_{\tilde \rho_{|\Omega}}. \]
  To upper bound $H_{\max}^{\frac{\epsilon}{4}}\big( \tilde C_1^n|  T_1^n E\big)_{\tilde \rho_{|\Omega}}$ we will use \cite[Lemma 28]{RMW16}. Indeed
   $H_{\max}^{\frac{\epsilon}{4}}\big( \tilde C_1^n|  T_1^n E\big)_{\tilde \rho_{|\Omega}}$ can be bounded   exactly in the same as in \cite[Lemma 28]{RMW16}, and leads to:
   \begin{align}
     H_{\max}^{\epsilon/4} (\tilde C_1^n|T_1^n E)_{\tilde \rho_{|\Omega}}  &\hspace{2mm}\leq \mu n + n(\alpha-1)\log^2(7)+\frac{\alpha}{\alpha-1}\log\left(\frac{1}{p_\Omega}\right)  - \frac{\log(1-\sqrt{1-(\epsilon/4)^2})}{\alpha-1}\\
    &\hspace{2mm}\leq \mu n + n(\alpha-1)\log^2(7)-\frac{\log(p_\Omega^2(1-\sqrt{1-(\epsilon/4)^2}))}{\alpha-1},
   \end{align}
   for $\alpha\in ]1,2]$.\\
   Taking $\alpha=1+\sqrt{\frac{-\log(p_\Omega^2(1-\sqrt{1-(\epsilon/4)^2}))}{n \log^2(7)}}$ gives us,
   \begin{align}
       H_{\max}^{\epsilon/4}(\tilde C_1^n|T_1^n E)_{\tilde \rho_{|\Omega}} &\leq \mu n +2 \sqrt{n} \log(7) \sqrt{-\log(p_\Omega^2(1-\sqrt{1-(\epsilon/4)^2}))}.\label{upBound_Hmax}
   \end{align}
   Putting eq.~\eqref{eq:EAT_bound},\eqref{eq:bound1} and \eqref{upBound_Hmax} together gives us,
   \begin{align}
     H_{\min}^\epsilon\big({A'}_1^n | X_1^n &{Y_{(1\ldots N-1)}}_1^n  T_1^n E\big)_{\tilde \rho_{|\Omega}}\geq \big(f(\hat q, p_{\rm opt})-\mu\big)\cdot n - \tilde v \sqrt{n} + 3\log(1-\sqrt{1-(\epsilon/4)^2}),
   \end{align}
   where $\tilde v= 2\big(\log(13)+ (\hat f'(p_{\rm opt})+1) \big)\sqrt{1-2\log(\epsilon \cdot p_{\Omega})}+2 \log(7) \sqrt{-\log(p_\Omega^2(1-\sqrt{1-(\epsilon/4)^2}))}$.

   Since $H_{\min}^\epsilon\big({A'}_1^n|X_1^n {Y_{(1\ldots N-1)}}_1^n  T_1^n E\big)_{\mathcal{M}_{\rm EC}(\tilde \rho)_{|\Omega}}= H_{\min}^\epsilon\big({A'}_1^n|X_1^n {Y_{(1\ldots N-1)}}_1^n  T_1^n E\big)_{\tilde \rho_{|\Omega}}$,
   we have,
   \begin{align}
     H_{\min}^\epsilon\big({A'}_1^n | X_1^n &{Y_{(1\ldots N-1)}}_1^n  T_1^n E\big)_{\mathcal{M}_{\rm EC}(\tilde \rho)_{|\Omega}}\geq \big(f(\hat q,p_{\rm opt})-\mu\big)\cdot n - \tilde v \sqrt{n} + 3\log(1-\sqrt{1-(\epsilon/4)^2}).
   \end{align}
   This bound holds for any $p_{\rm opt} \in ]\mu 3/4, \mu (1/2+1/2\sqrt{2})[$.
\end{proof}

In the following Lemma we link the result of the previous Lemma to the real state $\rho$ generated by the
protocol \ref{PtolA:DICKA}. Indeed, in the real state, the ``Bell violation''  is
not estimated directly,  but via the error corrections that might fail with some small probability.
We show that the real state of the protocol, when conditioned on the event
that \textit{Protocol \ref{PtolA:DICKA} does not abort {\bf and} the error corrections were successful},
possesses a linear amount on entropy.

\begin{Lmm}\label{Lmm:bound_entropy}
Let us call $\hat \Omega$ the event of not aborting the Protocol \ref{PtolA:DICKA} and $\hat \Omega'$ the event $\hat \Omega$ \textbf{and} all the error correction protocols were successful, meaning that
$\forall k\in[N-1],\ K_{B_{(k)}}={A'}_1^n$ and Alice's guess $G_{(1\dots N-1)}$ is correct. Then, for any $\epsilon_{\rm EA},\epsilon'_{\rm EC},\epsilon \in ]0,1[$, Protocol \ref{PtolA:DICKA} either aborts with a probability
$1-P(\hat \Omega)\geq 1- \big(1-2(N-1)\epsilon'_{\rm EC}\big)\epsilon_{\rm EA} $ ($\Leftrightarrow P(\hat \Omega') \leq \epsilon_{\rm EA}$) or\\
\begin{align}
  \begin{split}
   &H_{\min}^\epsilon\big({A'}_1^n| X_1^n {Y_{(1\ldots N-1)}}_1^n  T_1^n O_A O_{(1\ldots N-1)} E\big)_{\rho_{|\hat \Omega'}} \geq \\
  &\hspace{1cm}\max_{3/4 \leq \frac{p_{\rm opt}}{\mu} \leq 1/2+1/2\sqrt{2}} \hspace{-0.2cm}
  n \left( \big(f(\hat q,p_{\rm opt})-\mu\big) -   \frac{2\big(\log(13)+ (\hat f'(p_{\rm opt})+1) \big) \sqrt{1-2\log(\epsilon \cdot \epsilon_{\rm EA})}}{\sqrt{n}} \right) \\
   & \hspace{1cm} - \sqrt{n} \left(2 \log(7) \sqrt{-\log(\epsilon_{\rm EA}^2(1-\sqrt{1-(\epsilon/4)^2}))}\right) + 3\log(1-\sqrt{1-(\epsilon/4)^2})-{\rm leak_{EC}}(O_A) \\
   &\hspace{9.8cm}- \sum_{k=1}^{N-1} {\rm leak_{EC}}(O_{(k)}),
 \end{split}
\end{align}\\
where  $\hat q=(\mu \delta,\mu-\mu \delta,1-\mu)^t$.
\end{Lmm}
\begin{proof}
Using the chain rule \cite[Lemma 6.8]{T15} we get:
\begin{align}
  H_{\min}^\epsilon\big({A'}_1^n|& X_1^n {Y_{(1\ldots N-1)}}_1^n  T_1^n O_A O_{(1\ldots N-1)} E\big)_{\rho_{|\hat \Omega'}} \notag\\
  &\geq H_{\min}^\epsilon\big({A'}_1^n|X_1^n {Y_{(1\ldots N-1)}}_1^n  T_1^n E\big)_{\rho_{|\hat \Omega'}} -{\rm leak_{EC}}(O_A) - \sum_{k=1}^{N-1} {\rm leak_{EC}}(O_{(k)}),
\end{align}
where ${\rm leak_{EC}}(O_A)$ is the leakage due to the error correction protocol (when the Bobs try to guess Alice's bits) and ${\rm leak_{EC}}(O_{(k)})$ is the leakage due to error
correction (when Alice tries to guess Bob$_k$'s test rounds bits). These leakages will be estimated in Section~\ref{Sec.keyrate}.

We now need to bound $H_{\min}^\epsilon\big({A'}_1^n|X_1^n {Y_{(1\ldots N-1)}}_1^n  T_1^n E\big)_{\rho_{|\hat \Omega'}}$.
Note that the reduced state on ${A'}_1^n  X_1^n {Y_{(1\ldots N-1)}}_1^n  T_1^n E$ of the global state at the end of the Protocol \ref{PtolA:DICKA} conditioned on the event $\hat \Omega'$ of not aborting
\emph{and} all the error correction protocol were successful, is equal to the state $\mathcal{M}_{\rm EC}(\tilde \rho_{{A'}_1^n  X_1^n {Y_{(1\ldots N-1)}}_1^n  T_1^n E})_{|\Omega}$, therefore using Lemma \ref{Lemma:EAT_ECsucc}
we get:
\begin{align}
  H_{\min}^\epsilon\big({A'}_1^n | X_1^n &{Y_{(1\ldots N-1)}}_1^n  T_1^n E\big)_{ \rho_{|\hat \Omega'}}\geq \big(f(\hat q,p_{\rm opt})-\mu\big)\cdot n - \tilde v \sqrt{n} + 3\log(1-\sqrt{1-(\epsilon/4)^2}), \label{eq:bound_min_entrop}
\end{align}
where $\tilde v= 2\big(\log(13)+ (\hat f'(p_{\rm opt})+1) \big)\sqrt{1-2\log(\epsilon \cdot p_{\hat \Omega'})}+2 \log(7) \sqrt{-\log(p_{\hat \Omega'}^2(1-\sqrt{1-(\epsilon/4)^2}))}$.
\end{proof}

The following Lemma concludes on the soundness of Protocol \ref{PtolA:DICKA}. To
do so we need to relate the event $\hat \Omega$ that
the protocol \ref{PtolA:DICKA} does not abort, with the event $\hat \Omega'$ that
protocol \ref{PtolA:DICKA} does not abort {\bf and} that the error corrections are
successful.

\begin{Lmm}\label{Lmm:soundnes}
  For any implementation of the Protocol \ref{PtolA:DICKA}, either the protocol  aborts with a probability greater than $1- \epsilon_{\rm EA}$ or it is $((N-1)\epsilon_{\rm EC}+\epsilon_{\rm PA}+\epsilon)$-correct-and-secret while producing
  keys of length $l$ defined in eq.~\eqref{eq:key_length}.
\end{Lmm}
\begin{proof}
  Let $\hat \Omega$ be the event of not aborting in the protocol \ref{PtolA:DICKA}, and $\hat \Omega'$ the event $\hat \Omega$ \textbf{and} all the error correction protocols were successful.
  According to Lemma \ref{Lmm:bound_entropy} we are into one of the two following cases:
  \begin{itemize}
    \item The protocol aborts with a probability $1-P(\hat \Omega) \geq 1- \big(1-2(N-1)\epsilon'_{\rm EC}\big)\epsilon_{\rm EA}$. This is equivalent to $P(\hat \Omega')\leq \epsilon_{\rm EA}$ and implies that $1-P(\hat \Omega)\geq 1-\epsilon_{\rm EA}$.
    \item The aborting probability is $1-P(\hat \Omega) \leq 1-\epsilon_{\rm EA}$ (which implies that $P(\hat \Omega')\geq \epsilon_{\rm EA}$) and the smooth min-entropy of the final state conditioned on $\hat \Omega'$ is bounded as in eq.~\eqref{eq:bound_min_entrop}.
    Conditioned on $\hat \Omega$ there is two cases:
    \begin{itemize}
      \item The error correction step failed. This happens with probability
      at most $2(N-1)\epsilon'_{\rm EC}$.
      \item The error correction were successful and then all the keys agree. We have then the event $\hat \Omega'$. Therefore according to Lemma \ref{Lmm:bound_entropy}
      the entropy is high enough to produce keys of length $l$ such that:
      \begin{align}
        \left\|\rho_{K_A E {|\hat \Omega}} - \frac{\id_A}{2^l}\otimes \rho_{E {|\hat \Omega}}\right\|_{\tr} \leq \epsilon_{\rm PA} + 2\epsilon,
      \end{align}
     where $\epsilon_{\rm PA}$ is the privacy amplification error probability and $\epsilon$ is the smoothing parameter.
    \end{itemize}
    By combining the two above cases we have that the Protocol \ref{PtolA:DICKA} is $(\epsilon_{\rm PA} +2(N-1)\epsilon'_{\rm EC} + 2 \epsilon)$-correct-and-secret.
    \end{itemize}
\end{proof}

\subsection{Asymptotic key rate analysis}\label{Sec.keyrate}

In this section we evaluate the asymptotic key rate of the DICKA Protocol \ref{PtolA:DICKA} and compare it to the case where the parties perform $N-1$ DIQKD protocols in order to establish a common key.
In implementations where the efficiency of generation of GHZ states is comparable to the efficiency of the generation of EPR pairs a common key using a DICKA protocol can be, in principle, stablished in a much smaller number of rounds, however one need to analyse how the QBER and the leakages in the error correction protocol affects the key generation.

To analyse the key rate we need to evaluate the length $l$ of the final key produced by Protocol \ref{PtolA:DICKA}, Eq. \eqref{eq:key_length}, and compute the rate $r:=\frac{l}{\text{\#rounds}}$.
To achieve this, we need to estimate the leakage due to the error correction step. We use in our analysis an error correction protocol based on universal hashing \cite{BS94,RW05}.
The size of the leakage is taken to be the amount of correction information needed if the implementation
were honest, for some abort probability of the error correction protocol of at most $\epsilon_{\rm EC}$, and such that the guess (when not aborting) is correct with probability at least $1-\epsilon'_{\rm EC}$.
For a given honest implementation, this leakage can be bounded as follows \cite{RW05}:
\begin{align}
  {\rm leak(O_A)} &\leq \max_{k\in [N-1]} H_0^{\tilde \epsilon_{\rm EC}}(A_1^n | {{B'}_{(k)}}_1^n X_1^n Y_{(1\ldots N-1)} T_1^n) + \log({\epsilon'_{\rm EC}}^{-1}),\\
  {\rm leak(O_{(k)})} &\leq H_0^{\tilde \epsilon_{\rm EC}}(B'_{(k), I} | {A'}_1^n X_1^n Y_{(1\ldots N-1)} T_1^n) + \log({\epsilon'_{\rm EC}}^{-1}),
\end{align}
for $\epsilon_{\rm EC} = \tilde \epsilon_{\rm EC} + \epsilon'_{\rm EC}$, $I:=\{i\in[n]: T_i=1\}$ and where $H_0^{\tilde \epsilon_{\rm EC}}$ is evaluated on the state produced by the honest implementation.
If it turns out that the implementation is not the expected one then the protocol will just abort with a higher probability but the security is not affected.

We will consider here one particular honest implementation to evaluate the leakage. Then we will compare it to what we would get using $N-1$ device independent quantum key distribution ($(N-1)\times$DIQKD) protocols to distribute the key to the $N$ parties. For the key rate of the latter we will use
the recent and most general analysis given in \cite{ADFORV18}.
 Of course the following calculations can be adapted to other implementations.\\

\begin{Lmm}[Asymptotic key rate]\label{Lemma:asympRate}
There exist an implementation of Protocol \ref{PtolA:DICKA} in which the achieved asymptotic key rate is given by
\begin{align}\label{rateCKAasym}
\begin{split}
  r_{N-{\rm CKA},\infty}&=1-h\left( \frac{1}{2}+ \frac{1}{2}\sqrt{16\left(\frac{\sqrt{1-2Q}^N}{2\sqrt{2}} + \frac{(1-2Q)\left(1-\sqrt{1-2Q}^{N-2}\right)}{4\sqrt{2}}\right)^2-1} \right)-h(Q),
  \end{split}
\end{align}
where $Q$ is the QBER between Alice and each of the Bobs.
\end{Lmm}

\begin{proof}
In the following analysis we chose an i.i.d.  honest implementation scenario where we assume that the channel between Alice and each of the Bobs is a depolarizing channel:
\begin{align}
  \mathcal{D}(\rho)= (1-p_{\rm dep}) \rho + p_{\rm dep} \frac{\id}{2},
\end{align}
for $p_{\rm dep} \in ]0,1[$. We will also apply this channel to model the noise on Alice's side. The state that is produced by Alice's source is supposed
to be an $N$-GHZ state denoted ${\rm GHZ}_N:=\ketbra{{\rm GHZ}_N}{{\rm GHZ}_N}$, where $\ket{{\rm GHZ}_N}:=\frac{\ket{0}^{\otimes N} + \ket{1}^{\otimes N}}{\sqrt{2}}$. Therefore the state shared between Alice and the Bobs in one round is
$\rho_{A B_{(1\ldots N-1)}}:=\mathcal{D}^{\otimes N} ({\rm GHZ}_N)$. The QBER between Alice and each of the Bobs can then be expressed as
$Q=\frac{2p_{\rm dep}-p_{\rm dep}^2}{2}$ ($\Leftrightarrow p_{\rm dep}=1-\sqrt{1- 2Q} $) and the expected winning probability
of the Parity-CHSH game is given by:
\begin{align} \label{pwin}
p_{\rm exp}= \left[\frac{1}{2} + \frac{(1-p_{\rm dep})^N }{2\sqrt{2}} + \frac{(1-p_{\rm dep})^2 (1-(1-p_{\rm dep})^{N-2})}{4\sqrt{2}}\right].
\end{align}

\begin{figure}[!h]
  \includegraphics[width=330pt]{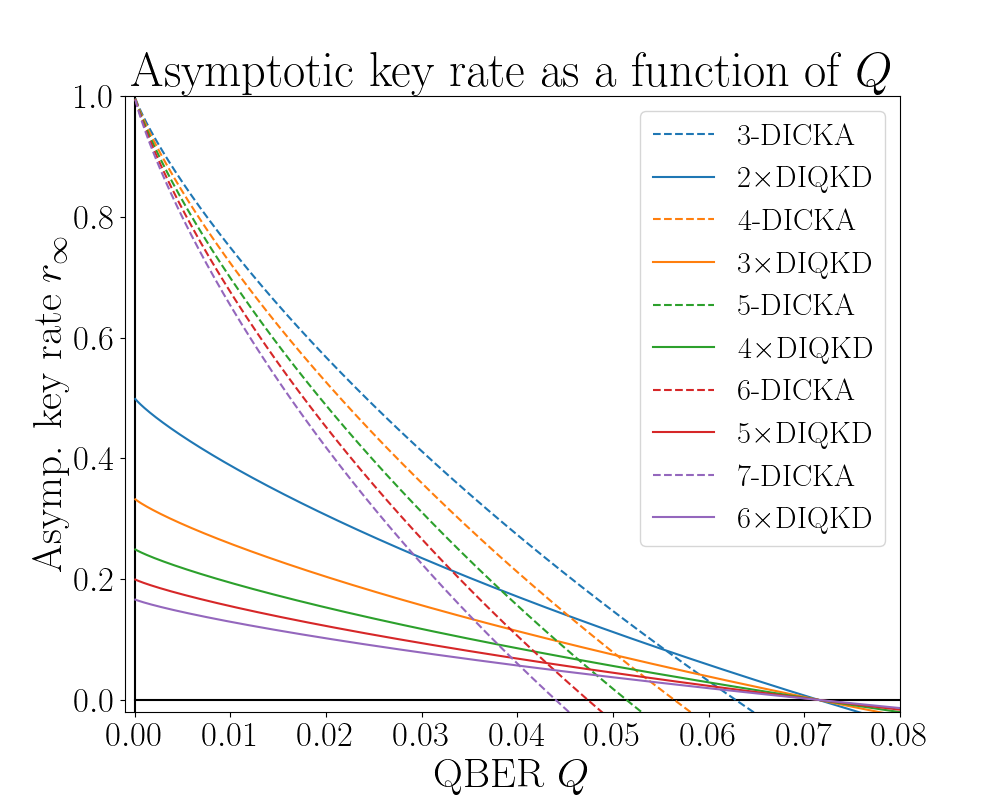}
  \caption{Asymptotic key rate for $N$-device independent CKA (DICKA, dashed lines), and for the distribution of a secret key between $N$ parties through $N-1$ device independent quantum key distribution ($(N-1)\times$DIQKD) protocols (solid lines), when each qubit experiences independent bit errors  measured at a bit error rate (QBER) $Q$.
  From top to  bottom, the lines correspond to $N=\{3,4,5,6,7\}$.
  We observe that for low noise it is advantageous to use our device independent $N$-CKA protocol instead of using $N-1$ DIQKD protocols \cite{ADFORV18}. In general, the comparison between the two methods depends on the cost and noisiness of producing GHZ states over pairwise EPR pairs.}
  \label{fig:asymp_key_rate}
\end{figure}

We can bound $H_0$ by $H_{\max}$ \cite[Lemma 18]{TSSR11} as,
\begin{align}
  H_0^{\tilde \epsilon_{\rm EC}}(A_1^n | {{B'}_{(k)}}_1^n X_1^n Y_{(1\ldots N-1)} T_1^n) \leq H_{\max}^{\tilde \epsilon_{\rm EC}/2}(A_1^n | {{B'}_{(k)}}_1^n X_1^n Y_{(1\ldots N-1)} T_1^n)+\log(8/{\tilde \epsilon_{\rm EC}^2} +2/(2-\tilde \epsilon_{\rm EC})).
\end{align}
Using the non-asymptotic version of the Asymptotic Equipartition Theorem \cite[Theorem 9]{TCR09} we get:
\begin{align}
  H_{\max}^{\tilde \epsilon_{\rm EC}/2}(A_1^n | {{B'}_{(k)}}_1^n X_1^n Y_{(1\ldots N-1)} T_1^n) \leq n H(A'_i| B'_{(1\ldots N-1),i} X_i Y_{(1\ldots N-1),i} T_i) + \sqrt{n} \Delta(\tilde \epsilon_{\rm EC}),
\end{align}
where $\Delta(\tilde \epsilon_{\rm EC}):= 4 \log\Big(2\sqrt{2^{H_{\max}(A'_i| B'_{(1\ldots N-1),i} X_i Y_{(1\ldots N-1),i} T_i)}}+1\Big) \cdot \sqrt{2 \log({8/{\tilde \epsilon_{\rm EC}^2}})}$.
We can now upper bound the entropy for honest implementation of Protocol \ref{PtolA:DICKA} as,
\begin{align}
  H(A'_i| B'_{(1\ldots N-1),i} X_i Y_{(1\ldots N-1),i} T_i)&= (1-\mu)\cdot H(A'_i| B'_{(1\ldots N-1),i} X_i Y_{(1\ldots N-1),i}, T_i=0) \\
  &\hspace{2cm} \notag +\mu \cdot \underbrace{H(A'_i| B'_{(1\ldots N-1),i} X_i Y_{(1\ldots N-1),i}, T_i=1)}_{\leq 1}\\
  &\leq (1-\mu) \cdot h(Q)+ \mu,
\end{align}
and $H_{\max}(A'_i| B'_{(1\ldots N-1),i} X_i Y_{(1\ldots N-1),i} T_i) \leq 1$.
This gives us an upper bound on ${\rm leak}(O_A)$:
\begin{align}
  \begin{split}
  &{\rm leak}(O_{A}) \\
  &\hspace{5mm}\leq n \cdot \big((1-\mu) \cdot h(Q)+ \mu\big) + \sqrt{n} \cdot 4 \log\Big(2\sqrt{2}+1\Big) \cdot \sqrt{2 \log({8/{\tilde \epsilon_{\rm EC}^2}})} +\log(8/{\tilde \epsilon_{\rm EC}^2} +2/(2-\tilde \epsilon_{\rm EC})).
\end{split}
\end{align}
Using the same reasoning, we get:
\begin{align}
  {\rm leak}(O_{(k)}) \leq n \cdot \mu + \sqrt{n} \cdot 4 \log\Big(2\sqrt{2}+1\Big) \cdot \sqrt{2 \log({8/{\tilde \epsilon_{\rm EC}^2}})} +\log(8/{\tilde \epsilon_{\rm EC}^2} +2/(2-\tilde \epsilon_{\rm EC})).
\end{align}

Putting this into equation \eqref{eq:key_length} we get,
\begin{align}
  l=&   \big(f(\hat q,p_{\rm opt})-(1-\mu) h(Q)- (N+1)\mu\big)\cdot n - \hat v \sqrt{n} + 3\log(1-\sqrt{1-(\epsilon/4)^2})-\log(\epsilon_{\rm PA}^{-1}) \notag\\
  & \hspace{8cm}-N \cdot \log(8/{\tilde \epsilon_{\rm EC}^2} +2/(2-\tilde \epsilon_{\rm EC})),
\end{align}
where $\hat v= \tilde v + N\cdot 4 \log\Big(2\sqrt{2}+1\Big) \cdot \sqrt{2 \log({8/{\tilde \epsilon_{\rm EC}^2}})}$, and $\tilde v$ is defined in Theorem \ref{Thm:Security}.

Note that in the asymptotic regime $n \to \infty$ we can take the threshold $\delta$ to be $\delta = p_{\rm exp}$, and the optimal $p_{\rm opt}$ will be
$p_{\rm opt}=\mu \delta = \mu p_{\rm exp}$. Also for the asymptotic analysis we chose $\mu=n^{-1/10}$. Therefore the asymptotic rate $r_{\infty}:= \lim_{n \to \infty} \frac{l}{\# {\rm rounds}}$ becomes,
\begin{align}\label{rateCKA}
\begin{split}
  \hspace{-0.45cm}r_{N-{\rm CKA},\infty}&=\hat f(\mu p_{\rm exp}) - h(Q)\\
  &=1-h\left( \frac{1}{2}+ \frac{1}{2}\sqrt{16\left(\frac{\sqrt{1-2Q}^N}{2\sqrt{2}} + \frac{(1-2Q)\left(1-\sqrt{1-2Q}^{N-2}\right)}{4\sqrt{2}}\right)^2-1} \right)-h(Q).
  \end{split}
\end{align}
\end{proof}

We then compare it to the asymptotic rate we would get if in order to distribute a key to $N$ parties, Alice were to use a DIQKD protocol for each of the Bobs.
To get the asymptotic rate for the $(N-1)$ DIQKD protocols, we use the analysis given in \cite{ADFORV18}. In their DIQKD protocol they consider an honest implementation
where the state is a depolarized EPR pair $(1-\nu) \Phi_{AB} + \nu \frac{\id}{2}$. If we say that, for each Bob, Alice sends the state via the same depolarizing
channel she uses in the previous analysis (and that she has the same noise on her qubits), we can link the parameter $\nu$ with the depolarizing parameter $p_{\rm dep}$ of the channel and to the QBER $Q$: $\nu = 2p-p^2 = 2 Q$.
Therefore we get:
\begin{align}\label{rateNDIQKD}
  r_{(N-1)\times{\rm QKD},\infty}= \frac{1-h\left( \frac{1}{2}+ \frac{1}{2}\sqrt{{2\cdot  (1- 2Q)^2 }-1} \right)-h(Q)}{N-1}.
\end{align}
Note that the factor $1/(N-1)$ comes from the fact that the total number of rounds while running $N-1$ DIQKD protocols is
$(N-1)n$, where $n$ is the number of rounds for one DIQKD protocol.

The comparison of the key rates of DICKA, Eq.~\eqref{rateCKA}, and $(N-1)\times$DIQKD, Eq.~\eqref{rateNDIQKD}, for different values of $N$, are plotted in Fig.~\ref{fig:asymp_key_rate}. The results show that for low noise it is advantageous to use the DICKA protocol.
In this comparison we assume  that the cost of generation of a GHZ state is the same as the cost to generate one EPR pair. However, in implementations where the GHZ state is created out of EPR pairs that will not be the case. Therefore the cost of creation of these states must be taken into account in the analysis of the  particular implementations.
Note, also, that in this Section we have modelled  the implementation for depolarising channels, however the security analysis is general and can be adapted for any particular implementation.

\end{widetext}

\bibliography{bib_article}
\bibliographystyle{unsrt}

\end{document}